\documentclass[]{emulateapj}
\usepackage{natbib}
\usepackage{amssymb}
\usepackage{color}
\usepackage{amsmath,mathtools}
\usepackage{epsfig}
\usepackage[FIGTOPCAP]{subfigure}
\usepackage{afterpage}
\usepackage{enumerate}
\usepackage{multirow}
\usepackage{verbatim}
\usepackage{tikz}
\usetikzlibrary{shapes.geometric, arrows}
\usepackage{morefloats}
\usepackage{wasysym}

\newcommand{\noop}[1]{}

\newcommand{\unit}[1]{\ensuremath{\, \mathrm{#1}}}

\shorttitle{Water Loss in the HZ of M Dwarfs}
\shortauthors{Luger \& Barnes 2014}

\begin{document}

\title{Extreme Water Loss and Abiotic O$_2$ Buildup \\On Planets Throughout the Habitable Zones of M Dwarfs}

\author{R. Luger\altaffilmark{1,2} and R. Barnes\altaffilmark{1,2}}
\affil{\altaffilmark{1}Astronomy Department, University of Washington, Box 351580, Seattle, WA 98195, USA; rodluger@uw.edu\\
\altaffilmark{2}Virtual Planetary Laboratory, Seattle, WA 98195, USA}

\begin{abstract}
We show that terrestrial planets in the habitable zones of M dwarfs older than $\sim$ 1 Gyr could have been in runaway greenhouses for several hundred Myr following their formation due to the star's extended pre-main sequence phase, provided they form with abundant surface water. Such prolonged runaway greenhouses can lead to planetary evolution divergent from that of Earth. During this early runaway phase, photolysis of water vapor and hydrogen/oxygen escape to space can lead to the loss of several Earth oceans of water from planets throughout the habitable zone, regardless of whether the escape is energy-limited or diffusion-limited. We find that the amount of water lost scales with the planet mass, since the diffusion-limited hydrogen escape flux is proportional to the planet surface gravity. In addition to undergoing potential desiccation, planets with inefficient oxygen sinks at the surface may build up hundreds to thousands of bars of abiotically produced O$_2$, resulting in potential false positives for life. The amount of O$_2$ that builds up also scales with the planet mass; we find that O$_2$ builds up at a constant rate that is controlled by diffusion: $\sim$ 5 bars/Myr on Earth-mass planets and up to $\sim$ 25 bars/Myr on super-Earths. As a result, some recently discovered super-Earths in the habitable zone such as GJ 667Cc could have built up as many as 2000 bars of O$_2$ due to the loss of up to 10 Earth oceans of water. The fate of a given planet strongly depends on the extreme ultraviolet flux, the duration of the runaway regime, the initial water content, and the rate at which oxygen is absorbed by the surface. In general, we find that the initial phase of high luminosity may compromise the habitability of many terrestrial planets orbiting low mass stars. \vspace*{0.3in}
\end{abstract}

\section{Introduction}
\label{sec:intro}
The discovery of the first Earth-sized planet in the habitable zone (HZ) of another star \citep{QUI14} marks the beginning of a new era in the study of exoplanets. With several upcoming missions capable of detecting potential Earth analogs around low mass stars, including TESS, K2, and PLATO \citep{RIC10,HOW14,RAU14}, it is imperative that we understand the processes that govern whether a planet in the HZ is in fact habitable. Given that over 40\% of M dwarfs (the lowest mass stars, spanning the range $0.08 \mathrm{M_\odot} \lesssim M \lesssim 0.6 \mathrm{M_\odot}$) are expected to harbor an Earth-sized planet in the HZ \citep{KOP13b}, the detailed spectroscopic characterization of all future detections may be very difficult. While next-generation space telescopes such as JWST may be capable of detecting certain biosignatures in these planets' atmospheres \citep[e.g.,][]{HED13,MIS14}, such observations will be extremely costly and require extensive amounts of valuable telescope time. Knowing in advance which planets are viable candidates for hosting life is therefore crucial, since it is possible that many planets in the HZ are not actually habitable for life as we know it. In particular, planets that form \emph{in situ} in the HZs of M dwarfs could be small and dry \citep{RAY07,LIS07}, while those that migrate from farther out could be unable to shed their thick H/He envelopes if they are more massive than about 1 M$_\oplus$ \citep{LAM14,LUG14}.

Moreover, planets around M dwarfs are subject to an array of processes that could negatively impact their habitability. M dwarfs are extremely active \citep{RH05,SCA07}, emitting large fractions of their luminosity in the X-ray and extreme ultraviolet (jointly referred to as XUV, corresponding to wavelengths of roughly 1-1000 $\mathrm{\AA}$). XUV photons are not only biologically harmful, but can drive fast atmospheric escape that leads to the erosion of planetary atmospheres \citep{WAT81,LAM03,YEL04,ERK07,TIA09,LAM09,OJ12,LAM13,ERK13,KOS13a,KOS13b}. Moreover, the HZs of these stars are significantly closer in, exposing planets to potentially catastrophic flaring events \citep{SEG10} and strong, detrimental tidal effects \citep{BAR13}.

However, an issue that is often overlooked is the fact that M dwarfs can take up to 1 Gyr to settle onto the main sequence (MS) because of their extended Kelvin-Helmholtz contraction timescales \citep[see, e.g.,][]{BAR98,RH05,DOT08}. During the contraction phase following their formation, these stars can be one or even two orders of magnitude more luminous than when they reach the MS. Since terrestrial planets probably form between 10 and 100 Myr after the formation of the star \citep{CHA04,RAY07,KLE09,RAY13}, planets in the HZs of these stars are subject to extreme levels of insolation early on and are likely to be in a runaway greenhouse provided they have sufficient surface water \citep[see, e.g.,][]{KAS88,KOP13}. 

Many papers have explored the effects of a runaway greenhouse on Venus, arguing that it may have lost one or more Earth oceans of water as a consequence of an early runaway \citep{KAS84,KAS88,CHA96,CHA96I,KUL06,GIL09}. During a runaway greenhouse, water vapor reaches the stratosphere, where it is easily photolyzed by UV radiation. Heating of the upper atmosphere by XUV radiation can then drive a hydrodynamic wind that carries the hydrogen (and potentially some of the oxygen) to space, leading to the irreversible loss of a planet's surface water, oxidation of the surface, and possible accumulation of oxygen in the atmosphere. Recently, \cite{HAM13} extended this idea to exoplanetary systems, arguing for the existence of two fundamentally different types of terrestrial planets: type I planets, which undergo short-lived runaway greenhouses during their formation, and type II planets, which form interior to a critical distance and can remain in runaway greenhouses for as long as 100 Myr. The former type of planet, like Earth, retains most of its water inventory and may evolve to become habitable. The latter, similarly to Venus, undergoes complete surface desiccation during the runaway.

In this paper we show that because of the early evolution of the star, many terrestrial planets within the HZ of M dwarfs could be similar to type II planets and may therefore be uninhabitable. Our work builds on that of \cite{BH13} and \cite{HB13}, who studied water loss during early runaway greenhouses on planets orbiting white dwarfs and brown dwarfs and on exomoons orbiting giant planets. We further build on the results of \cite{WP13}, who showed that significant water loss can occur for planets near the inner edge of the HZ of M dwarfs; however, those authors considered a constant stellar luminosity and thus did not account for the early runaway greenhouse state, which we show can result in water loss rates that are orders of magnitude higher. In a follow-up paper, \cite{WP14} showed that water loss can lead to the buildup of abiotic O$_2$ in the atmospheres of planets in the HZ. We extend this mechanism and demonstrate that hundreds to thousands of bars of abiotic oxygen are possible for planets throughout the HZs of M dwarfs. While a large fraction of this oxygen may be removed by surface processes, some exoplanets could retain detectable amounts of O$_2$ in their atmospheres for extended periods of time. This validates the predictions of \cite{SK00}, concerning oxygen atmospheres on Venus-like planets, and of \cite{LAM11a}, \cite{LAM11b}, \cite{LAM13BOOK}, and \cite{FOS14}, who argued that oxygen-rich atmospheres could develop on G dwarf planets in the HZ, in particular on super-Earths. Our present work could also strengthen the results of \cite{WP14} and \cite{TIA14} that oxygen is not a reliable biosignature; in fact, planets with such elevated quantities of O$_2$ may be uninhabitable.

The paper is organized as follows: in \S\ref{sec:moreintro} we describe the relevant physics, including the habitable zone limits, stellar evolution, hydrodynamic escape and oxygen buildup. In \S\ref{sec:model} we describe our model, followed by our results in \S\ref{sec:results1} and \S\ref{sec:results2}. We discuss the stability of O$_2$-rich atmospheres and implications for habitability in \S\ref{sec:discussion}.

\section{The Habitable Zone, Stellar Evolution, and Planetary Evolution}
\label{sec:moreintro}
\subsection{The Habitable Zone}
\label{sec:hz}
The habitable zone is the region around a star where liquid water may exist on the surface of a terrestrial planet \citep[e.g.,][]{KAS93,KOP13}. The outer edge of the HZ is traditionally given by the maximum greenhouse (MG) limit, beyond which the addition of CO$_2$ to the atmosphere is unable to provide sufficient greenhouse warming to prevent complete freezing of the oceans. The inner edge is the runaway greenhouse (RG) limit, interior to which a planet is unable to cool sufficiently to prevent the complete evaporation of its surface water. During a runaway greenhouse, rising temperatures result in net evaporation of the planet's oceans, which progressively increase the vapor pressure of the air up to a point where the atmosphere becomes optically thick in the infrared. The surface is then unable to cool effectively and the temperature increases to $\sim$ 1500 K, leading to the complete evaporation of the planet's oceans and effectively sterilizing the surface \citep[for a review, see][]{PIE10,GW12}.

It is important to note, however, that the MG and RG limits are by no means sharp edges to the HZ. Classical calculations of the HZ boundaries \citep{KAS88,KOP13,KOP14} rely on  one-dimensional models tailored to reproduce conditions on Earth and are unable to capture changes in the atmospheric circulation and cloud formation mechanisms that occur as a planet begins to warm. Recent studies such as those of \cite{ABE11}, \cite{LEC13}, and \cite{YAN14} used global climate models to show that the threshold for a runaway greenhouse can be significantly higher than that predicted by 1D models and may be quite sensitive to factors such as the planet's rotation rate and surface water content. Because of the uncertainties regarding the actual edges of the theoretical HZ, studies frequently define a wider ``empirical'' HZ based on evidence for the presence of water on Venus and Mars in the past \citep{KOP13}. The empirical HZ is bounded by the ``recent Venus'' (RV) limit at the inner edge and the ``early Mars'' limit at the outer edge. In general, we may expect the true HZ to lie somewhere in between the theoretical (RG and MG) and empirical (RV and EM) limits. 

We note, finally, that the location of the RG limit is a weak function of planet mass. This is due to the fact that the pressure at the emission level of a saturated atmosphere  scales with surface gravity; at a higher pressure, the temperature of the emission level is higher and the planet is able to cool more effectively, delaying the runaway state \citep{PIE10}.

\begin{figure*}[t]
  \begin{center}
    \leavevmode
      \psfig{file=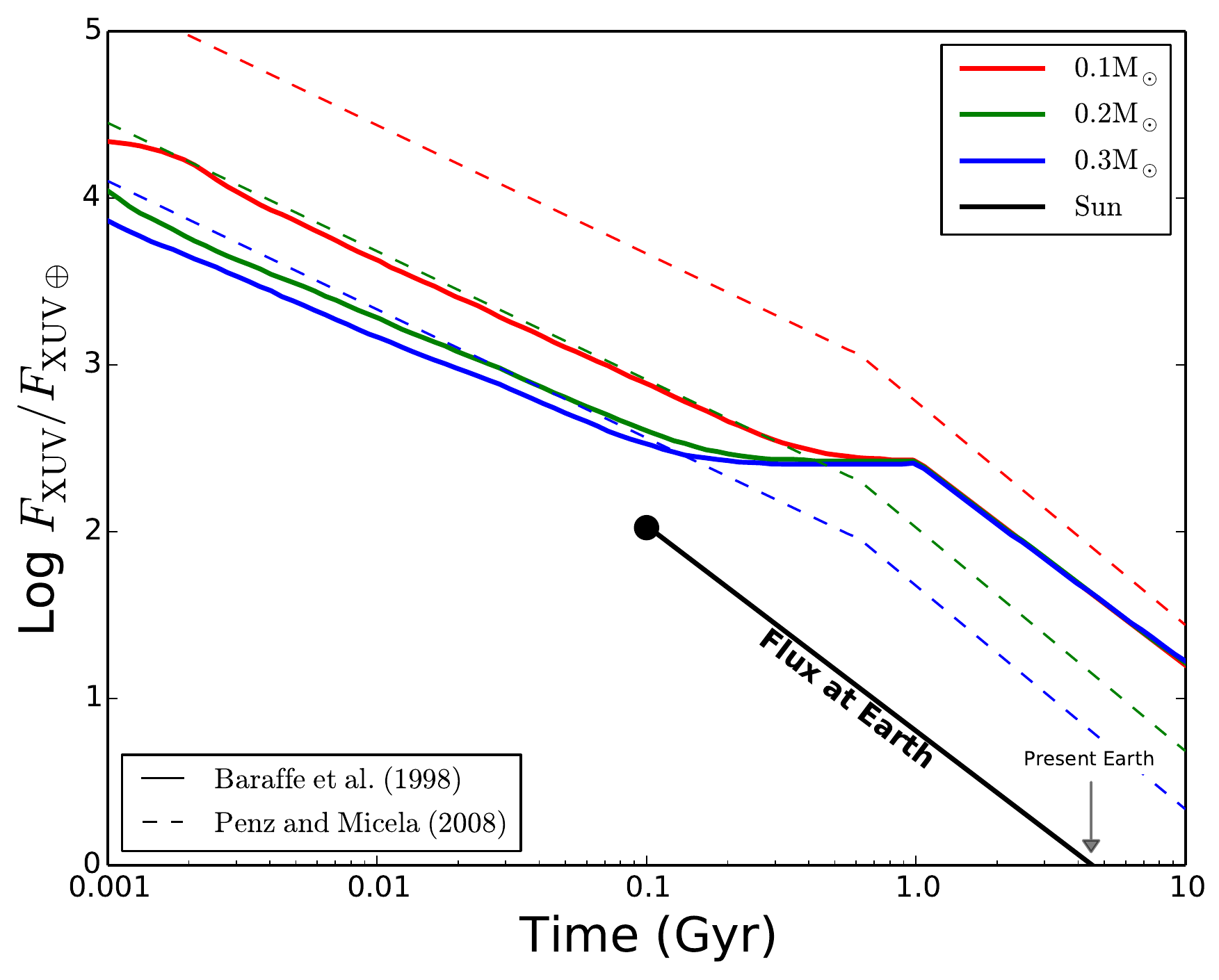,width=4in}
       \caption{Evolution of the XUV flux received by planets close to the inner edge of the HZ orbiting 0.1, 0.2 and 0.3$\mathrm{M_\odot}$ M dwarfs, scaled to that received by the present Earth, $\mathcal{F}_\oplus = 4.64\ \mathrm{erg/cm^2/s}$. Solid lines correspond to the model adopted in this paper, where fluxes are calculated using the stellar evolution model of \cite{BAR98} and the XUV evolution of \cite{RIB05}, assuming a saturation time $t_\mathrm{sat}$ of 1 Gyr and a saturation fraction of $10^{-3}$. Dashed lines correspond to the empirical model of \cite{PM08}, who assume a single XUV luminosity for all M dwarfs; see text for a discussion.
       The evolution of the flux at Earth is shown for reference as a black line. The dot corresponds to the earliest time for which the study of \cite{RIB05} has data for solar-type stars and is approximately equal to the saturation time for the Sun; it is also roughly the formation time for Earth.
       }
     \label{fig:xuvevol}
  \end{center}
\end{figure*}

\subsection{Stellar Evolution}
\label{sec:stellar}
\subsubsection{Bolometric Luminosity}
Following its formation from a giant molecular cloud, a low mass star will slowly contract under its own gravity along the Hayashi track \citep{HAY61} until its core temperature is high enough to ignite hydrogen fusion, at which point it is said to have reached the \emph{main sequence} (MS). The duration of the pre-main sequence (PMS) phase is inversely proportional to the mass of the star; while a star like the Sun reaches the MS in $\lesssim$ 50 Myr \citep{BAR98}, M dwarfs can take several hundred Myr to fully contract and reach the MS \citep{RH05}. During this time, the star's shrinking radius and roughly constant effective temperature result in a decrease in its luminosity by one or even two orders of magnitude. Because of this, the location of the HZ limits will vary significantly in the first few 100 Myr; planets in the HZ of M dwarfs today were probably not in the HZ when they formed.

Once a star reaches the MS, the steady increase in the mean atomic weight of its core due to fusion leads to a rise in its central temperature, thereby slowly increasing the luminosity. While the Sun brightens by roughly 10\% per Gyr, low mass M dwarfs fuse hydrogen very slowly once they are on the MS and can have nearly constant luminosities for tens of Gyr.

\subsubsection{XUV Luminosity}
The XUV luminosity of M dwarfs also evolves with time. Since XUV emission is linked to stellar magnetic activity, which declines with age, the XUV flux a planet receives will also decrease over time. For solar-type stars, \cite{RIB05} found that the evolution of the XUV luminosity follows a simple power law with an initial short-lived, constant ``saturation'' phase:
\begin{align}
\label{eq:lxuv}
\frac{L_\mathrm{XUV}}{L_\mathrm{bol}} = \left\{
				\begin{array}{lcr}
					f_0 &\ & t \leq t_0 \\
					f_0\left(\frac{t}{t_0}\right)^{-\beta} &\ & t > t_0,
				\end{array}
				\right.
\end{align}
where $L_\mathrm{bol}$ is the total (bolometric) stellar luminosity, $\beta = -1.23$, and $f_0$ is the initial (constant) ratio of XUV to bolometric luminosity. Prior to $t = t_0$, the XUV luminosity is said to be ``saturated,'' as observations show that the ratio $L_\mathrm{XUV}/L_\mathrm{bol}$ remains relatively constant at early times. \cite{JDW12} found that $t_\mathrm{sat} \approx 100 \unit{Myr}$ and $f_0 \approx 10^{-3}$ for K dwarfs (stars with 0.6 M$_\odot$ to 0.9 M$_\odot$). Given poor constraints on the ages of field M dwarfs, determining the exact value of the saturation timescale for these stars remains difficult. However, it is likely that their saturation timescale is much longer. \cite{WRI11} showed that X-ray emission from low mass stars is saturated for values of the Rossby number $Ro \equiv P_\mathrm{rot}/\tau \lesssim 0.1$, where $P_\mathrm{rot}$ is the stellar rotation period and $\tau$ is the convective turnover time. The extent of the convective zone increases with decreasing stellar mass; below 0.35M$_\odot$, M dwarfs are fully convective \citep{CB97}, resulting in larger values of $\tau$ \citep[see, e.g.,][]{PIZ00}. Low mass stars also have longer spin-down times \citep{STA94} and therefore smaller values of $Ro$ at a given age, leading to longer saturation times compared to solar-type stars. This is consistent with observational studies of late M dwarfs; in particular, \cite{WES08} showed that the magnetic activity lifetime increases from $\lesssim 1 \unit{Gyr}$ for early (i.e., most massive) M dwarfs to $\gtrsim 7 \unit{Gyr}$ for late (least massive) M dwarfs.

In Figure~\ref{fig:xuvevol} we plot the evolution of the XUV flux on planets close to the inner edge of the HZ for M dwarfs of mass 0.1, 0.2, and 0.3$\mathrm{M_\odot}$, assuming a saturation time $t_\mathrm{sat} = 1\ \mathrm{Gyr}$. The black line corresponds to the present-day XUV flux on Earth, $\mathcal{F}_\oplus = 4.64\ \mathrm{erg/cm^2/s}$ \citep{RIB05}. Note that the XUV flux around M dwarfs can be orders of magnitude higher than $1 \mathcal{F}_\oplus$. For reference, we plot the XUV fluxes calculated from the equations in \cite{PM08} and \cite{LAM09} as dashed lines. These studies derive an empirical power-law fit to the XUV luminosity of stars in the Pleiades and Hyades clusters; however, they present a single luminosity curve for all M dwarfs, which is an oversimplification that neglects the spread in bolometric luminosity over more than two orders of magnitude among M dwarfs. For intermediate mass M dwarfs, the fluxes in the HZ are somewhat comparable to those obtained from the tracks of \cite{BAR98}, but for the lowest mass M dwarfs the study of \cite{PM08} is likely to significantly overestimate the XUV flux.

\subsection{Planet Formation \& Initial Water Content}
\label{sec:planetformation}
The terrestrial planets in the solar system are thought to have formed \emph{in situ} between 10 and 100 Myr after the formation of the Sun \citep{CHA04,KLE09,RAY13}. However, whether or not the bulk of Earth's oceans formed during this accretion period is still up to debate. A recent isotopic study by \cite{HAR11} suggests that a large fraction of Earth's water may have been delivered by comets, possibly a result of scattering by the giant planets prior to and during the Late Heavy Bombardment \citep{GOM05}. Nevertheless, simulations show that during the final stages of Earth's assembly, the planet's feeding zone encompassed enough water-rich material to supply $15-70$ terrestrial oceans (TO) of water in the first 70 Myr \citep{MOR00,RAY06,CHA12}, and thus a wet \emph{in situ} formation for Earth-like planets is entirely plausible. In this paper, we define 1 TO $\equiv 1.39\times 10^{24}$ g ($\sim 270$ bars) of H$_2$O, the total amount of water in Earth's oceans \citep{KAS88,KUL06}.

Whether or not this applies to M dwarfs is unclear. Studies by \cite{RAY07} and \cite{LIS07} suggest that planets forming \emph{in situ} in the HZs of M dwarfs are likely to form quickly ($\sim 10$ Myr after the formation of the star), to be relatively small ($\lesssim 0.3\mathrm{M_\oplus}$) and to have water contents smaller than Earth's. If that is the case, a potential mechanism for forming a wet, Earth-size planet in the HZ is early formation beyond the snow line (the region of the circumstellar disk beyond which water and other volatiles are able to condense into ices and serve as building blocks for planets) followed by disk-driven migration into the HZ.

Planets that form prior to the dissipation of the gaseous circumstellar disk experience strong torques that can induce rapid inward migration, especially for planets in the terrestrial mass range \citep{WAR97}. In particular, planets that form beyond the snow line, where accretion is orders of magnitude faster due to the higher density of solids, can potentially migrate into the HZ \citep{IL08a, IL08b, OI09, COS13}. Since disk lifetimes are typically quite short, ranging from $\sim 1$ to $\sim 10$ Myr \citep{WAL88,STR89}, planets migrating in this fashion will settle into their new orbits relatively early. As \cite{RAY14} show, the abundance of short-period planets with masses $\lesssim 10$ M$_\oplus$ is strong evidence for the ubiquity of this mechanism, since it is highly unlikely that these systems formed \emph{in situ}. Because of their formation beyond the snow line, these planets will likely have large ice mass fractions and therefore much larger initial water contents than Earth \citep[see, e.g.,][]{KUC03}.

It is important to note that not all of a planet's water may be at its surface (or in its atmosphere), particularly at early times. During terrestrial planet formation, giant impacts can deliver enough energy to partially or completely melt a planet's mantle; as a consequence, many of the terrestrial bodies in the solar system may have experienced a magma ocean phase \citep{MA86,ZAH88,ES08b,ELK08,ELK11,ELK12,LAM13BOOK,LEB13,HAM13}. Since water is highly soluble in magma, a large fraction of the planet's water content will initially be trapped in the mantle. As the planet cools and the mantle begins to solidify from the bottom up, large amounts of water (between $\sim$ 60 and 99\% of the total amount in the mantle) are exsolved to form a massive steam atmosphere, which may eventually collapse to form an ocean \citep{ELK11}. Typically, this process occurs within a few Myr of the end of the accretion phase \citep{ELK08,ELK11}, but the exact timescale for solidification depends on the stellar flux. \cite{LEB13} find that while Earth's magma ocean lasted for $\sim$ 1.5 Myr, it may have lasted as long as 10 Myr on Venus due to the blanketing effect of a runaway greenhouse. Furthermore, \cite{HAM13} argue that above a certain stellar flux (close to that received by Venus), a magma ocean may take as long as 100 Myr to solidify. While these planets never develop massive steam atmospheres---since the bulk of the water is always in the mantle---a few tens of bars of water vapor are always maintained in the atmosphere due to a feedback cycle \citep{MA86,ZAH88}. Large quantities of water may thus still be lost via hydrodynamic escape from these planets, since escape of water to space will be balanced by exsolution from the magma ocean. As \cite{HAM13} point out, this could lead to the complete desiccation of a planet's mantle, potentially terminating tectonics and resulting in permanently dry surface conditions.

One might thus expect that because of the high-luminosity PMS phase of M dwarfs, planets in the HZs of these stars could remain in the magma ocean phase for several to several tens of Myr, though this should be investigated further. While the magma ocean phase does not prevent water loss to space, it could suppress the buildup of atmospheric O$_2$. We revisit this point in \S\ref{sec:magmaocean}.

\subsection{Atmospheric Escape}
\label{sec:waterloss}
\subsubsection{Energy-Limited Escape}
During a runaway greenhouse, the surface temperature of a terrestrial planet exceeds the temperature at the critical point of water (647 K) and the oceans fully evaporate \citep{KAS88}. The mixing ratio of water vapor in the stratosphere (i.e., the ratio of the molar abundance of water to that of the background gas) approaches unity, and water molecules are exposed to high levels of XUV and far-UV (FUV) radiation, which photolyze the water, releasing hydrogen and oxygen. In the classical picture, hydrogen escapes to space while oxygen interacts with the surface, oxidizing the rocks. This is widely believed to have happened on Venus \citep{WAT81,CHA96}.

Under the high XUV irradiation of M dwarfs, hydrogen escape is thought to occur via a hydrodynamic wind and can be ``energy-limited,'' where a fixed fraction $\epsilon_\mathrm{XUV}$ of the incoming XUV energy goes into driving the escape
\citep{WAT81,ERK07,LAM13,VJ13,JVE13}. The energy-limited mass loss rate $\dot{M}_\mathrm{EL}$ is obtained by equating the energy provided by XUV photons to the energy required to lift the atmosphere out of the gravitational potential well. It may be written as \citep{ERK07}
\begin{align}
\dot{M}_\mathrm{EL} = \frac{\epsilon_\mathrm{XUV}\pi \mathcal{F}_\mathrm{XUV}R_\mathrm{p}R_\mathrm{XUV}^2}{GM_\mathrm{p}K_\mathrm{tide}}
\label{eq:dmdt}
\end{align}
where $\mathcal{F}_\mathrm{XUV}$ is the XUV flux, $M_\mathrm{p}$ is the mass of the planet, $R_\mathrm{p}$ is the planet radius, $R_\mathrm{XUV}$ is the radius where the bulk of the energy is deposited (which, for simplicity, we take to be equal to $R_\mathrm{p}$), $\epsilon_\mathrm{XUV}$ is the XUV absorption efficiency, and $K_\mathrm{tide}$ is a tidal correction term of order unity.

\subsubsection{Oxygen Escape}
\label{sec:oxygenescape}
Strong hydrodynamic flows are capable of dragging heavier species along with them. Given the high XUV fluxes of M dwarfs early on, it is important to consider the case where the oxygen escapes along with the hydrogen. \cite{HUN87} studied mass fractionation during hydrodynamic escape, demonstrating that an escaping species can efficiently drag a heavier species along with it provided the mass of the latter is smaller than the \emph{crossover mass} $m_\mathrm{c}$, equal to
\begin{align}
\label{eq:crossover}
m_\mathrm{c} = m_\mathrm{H} + \frac{kTF_\mathrm{H}}{bgX_\mathrm{H}},
\end{align}
in the case of a background flow of atomic hydrogen. Here, $m_\mathrm{H}$ is the mass of a hydrogen atom, $T$ is the temperature of the flow, $F_\mathrm{H}$ is the planet-averaged upward H particle flux, $b$ is the binary diffusion coefficient for the two species, $g$ is the acceleration due to gravity, and $X_\mathrm{H}$ is the H molar mixing ratio at the base of the flow. The oxygen particle flux $F_\mathrm{O}$ is then given by \citep{HUN87,CHA96I,LAM11a,ERK14}
\begin{align}
\label{eq:fo}
F_\mathrm{O} &= \frac{X_\mathrm{O}}{X_\mathrm{H}}F_\mathrm{H}\left( \frac{m_\mathrm{c}-m_\mathrm{O}}{m_\mathrm{c}-m_\mathrm{H}} \right),
\end{align}
where $X_\mathrm{O}$ is the oxygen mixing ratio and $m_\mathrm{O}$ is the mass of an oxygen atom. The expression above is valid provided $m_\mathrm{c} \geq m_\mathrm{O}$; otherwise, $F_\mathrm{O} = 0$. In the limit of large $m_\mathrm{c}$, the ratio of the particle escape rates is simply the ratio of the abundances at the base of the flow.

It is important to note that both $F_\mathrm{H}$ and $m_\mathrm{c}$ are indirect functions of $F_\mathrm{O}$; as oxygen begins to escape, the hydrogen particle flux decreases (at fixed energy input), decreasing the crossover mass and in turn reducing the rate of oxygen escape. In order to solve for the individual escape rates, it is convenient to define a reference particle flux $F_\mathrm{H}^\mathrm{ref}$, equal to the energy-limited particle escape flux of H in the absence of oxygen \citep{CHA96I}:
\begin{align}
\label{eq:fhref}
F_\mathrm{H}^\mathrm{ref} = \frac{\epsilon_\mathrm{XUV}\mathcal{F}_\mathrm{XUV}R_\mathrm{p}}{4GM_\mathrm{p}K_\mathrm{tide}m_\mathrm{H}},
\end{align}
where we may write
\begin{align}
\label{eq:melfhref}
m_\mathrm{H}F_\mathrm{H}^\mathrm{ref} = \frac{\dot{M}_\mathrm{EL}}{4\pi R_\mathrm{p}^2} = m_\mathrm{O}F_\mathrm{O} + m_\mathrm{H}F_\mathrm{H}.
\end{align}
By combining (\ref{eq:fo}) and (\ref{eq:melfhref}), we obtain as in \cite{CHA96I} the true hydrogen particle flux in terms of the reference flux:
\begin{align}
\label{eq:fhfhref}
F_\mathrm{H} =
  \begin{dcases}
   F_\mathrm{H}^\mathrm{ref} & \text{if } m_\mathrm{c} < m_\mathrm{O}\\
   F_\mathrm{H}^\mathrm{ref}\left( 1 + \frac{X_\mathrm{O}}{X_\mathrm{H}}\frac{m_\mathrm{O}}{m_\mathrm{H}}\frac{m_\mathrm{c}-m_\mathrm{O}}{m_\mathrm{c}-m_\mathrm{H}} \right)^{-1} & \text{if } m_\mathrm{c} \ge m_\mathrm{O}
  \end{dcases}
\end{align}
Inserting this into (\ref{eq:crossover}) and doing a little algebra, we obtain an expression for the crossover mass in terms of the reference flux (\ref{eq:fhref}):
\begin{align}
\label{eq:mcfhref}
m_\mathrm{c} =
  \begin{dcases}
   m_\mathrm{H} + \frac{3kTF_\mathrm{H}^\mathrm{ref}}{2bg} & \text{if } F_\mathrm{H}^\mathrm{ref} < 10bgm_\mathrm{H}/kT \\
   \frac{43}{3}m_\mathrm{H} + \frac{kTF_\mathrm{H}^\mathrm{ref}}{6bg} & \text{if } F_\mathrm{H}^\mathrm{ref} \ge  10bgm_\mathrm{H}/kT
  \end{dcases}
\end{align}
In this derivation, we used $m_\mathrm{O} = 16m_\mathrm{H}$ as well as $X_\mathrm{H} = 2/3$ and $X_\mathrm{O} = 1/3$, assuming that all H and O are photolytically produced and that the dissociation of H$_2$ and O$_2$ is fast enough that both species are atomic close to the base of the flow \citep{CHA96I}.

The condition for oxygen escape is $m_\mathrm{c} > m_\mathrm{O}$; using (\ref{eq:melfhref}), we can write this as $\mathcal{F}_\mathrm{XUV} \geq \mathcal{F}_\mathrm{crit}$, where
\begin{align}
\label{eq:fxuvmin}
\mathcal{F}_\mathrm{crit} \equiv 180 \left( \frac{M_\mathrm{p}}{\mathrm{M}_\oplus} \right)^{2} \left( \frac{R_\mathrm{p}}{\mathrm{R}_\oplus} \right)^{-3} \left(\frac{\epsilon_\mathrm{XUV}}{0.30}\right)^{-1}\ \mathrm{erg\ cm^{-2}\ s^{-1}}
\end{align}
where we have used $b = 4.8\times 10^{17}(T/\mathrm{K})^{0.75}\ \mathrm{cm^{-1}s^{-1}}$ \citep{ZK86}, an average thermospheric temperature $T = 400$ K \citep{HUN87,CHA96I}, and $K_\mathrm{tide} = 1$, corresponding to no tidal enhancement. For reference, a 1 M$_\oplus$, 1 R$_\oplus$ planet at the RV limit of a 0.1 M$_\odot$ star (for which tidal effects are strongest), has $K_\mathrm{tide} \approx 0.88$; setting $K_\mathrm{tide} = 1$ leads to a maximum underestimate of the mass loss rate of about 10\%. 

Thus, for an Earth-size terrestrial planet with $\mathcal{F}_\mathrm{XUV}~>~180\ \mathrm{erg\ cm^{-2}\ s^{-1}}$ (equivalent to $\sim39\mathcal{F}_\oplus$), oxygen should begin to escape. The particle escape flux is determined by inserting (\ref{eq:mcfhref}) into (\ref{eq:fo}):
\begin{align}
\label{eq:oxygenflux}
F_\mathrm{O} &= \frac{\eta}{2}F_\mathrm{H},
\end{align}
where we define the oxygen escape parameter $\eta$ as
\begin{align}
\label{eq:eta}
\eta \equiv
  \begin{dcases}
   0       & \text{if } x < 1 \\
   \frac{x-1}{x+8} & \text{if } x \geq 1
  \end{dcases}
\end{align}
with
\begin{align}
\label{eq:x}
x \equiv \frac{kTF_\mathrm{H}^\mathrm{ref}}{10bgm_\mathrm{H}}.
\end{align}
The parameter $\eta$ is simply the ratio of the oxygen particle flux to its production rate (compare Equations \ref{eq:simple8} and \ref{eq:moupmhup} in the Appendix). It is thus limited to the range $0 \leq \eta \leq 1$. For $\eta \rightarrow 1$ (large $x$, high $\mathcal{F}_\mathrm{XUV}$), the oxygen flux approaches one-half the hydrogen flux, resulting in no accumulation of oxygen in the atmosphere. For $\eta \rightarrow 0$ ($x \rightarrow 1$, low $\mathcal{F}_\mathrm{XUV}$), oxygen escape tapers off and is zero for $x < 1$. Note that the condition $x \geq 1$ is mathematically equivalent to both $\mathcal{F}_\mathrm{XUV} \geq \mathcal{F}_\mathrm{crit}$ and $m_\mathrm{c} \geq m_\mathrm{O}$ in (\ref{eq:mcfhref}). In Appendix A, we derive simple expressions for the mass loss rates of hydrogen (\ref{eq:mdothup}) and oxygen (\ref{eq:mdotoup}), the rate of oxygen buildup (\ref{eq:mdotodown}), and the rate of ocean loss (\ref{eq:mdotocean}) in terms of $\eta$ and $\dot{M}_\mathrm{EL}$.

\subsubsection{Diffusion Through an Oxygen Atmosphere}
\label{sec:difflim}
The energy-limited mass loss rate (\ref{eq:dmdt}) is an upper limit to the thermal escape rate from a planetary atmosphere, as it assumes that all of the available XUV energy goes into driving the escape (after accounting for an efficiency factor, $\epsilon_\mathrm{XUV}$). However, the escape of hydrogen depends on the availability of hydrogen atoms at the base of the flow. If not all of the oxygen escapes or is absorbed by the surface, it can eventually become a major constituent of the atmosphere. Once this happens, hydrogen will have to diffuse through a static background of oxygen before it reaches the base of the hydrodynamic wind. The rate at which hydrogen can do so is given by the \emph{diffusion limit} \citep{HUN73,ZKP90}:
\begin{align}
\label{eq:difflim}
F_\mathrm{H}^\mathrm{diff} &\equiv \frac{bg(m_\mathrm{O}-m_\mathrm{H})}{kT(1+X_\mathrm{O}/X_\mathrm{H})},
\end{align}
which can be significantly lower than energy-limited escape flux (\ref{eq:fhfhref}), especially at early times.\footnote{In (\ref{eq:difflim}), we (conservatively) assumed diffusion through atomic oxygen. Diffusion through molecular oxygen is faster, but if the oxygen is photolyzed below the base of the hydrodynamic wind, it is diffusion through atomic oxygen that will bottleneck the escape.} In particular, note that when $X_\mathrm{O}/X_\mathrm{H} = 1/2$, $F_\mathrm{H}^\mathrm{diff} = 10bgm_\mathrm{H}/kT$, which is precisely the value of the reference flux at which oxygen begins to escape in (\ref{eq:mcfhref}). The diffusion limit is, in fact, defined as the maximum upward flux of a gas for which the background gas is static \citep{HUN73}. If the hydrogen particle flux is greater than this limit, oxygen \emph{must} escape. The converse is also true; if hydrogen diffusion through oxygen in the lower atmosphere limits the supply of H atoms at the base of the flow, the escaping flux will be insufficient to drag away any oxygen.

Thus, for planets that build up significant amounts of oxygen in their atmospheres, the H particle escape flux is given by the smaller of (\ref{eq:fhfhref}) and (\ref{eq:difflim}), the O particle escape flux is zero, and the rate at which the ocean is lost is 9 times the H escape rate.

\subsection{Oxygen Atmospheres}
\label{sec:o2atmospheres}

\subsubsection{Earth-like Planets}
Because of its highly reactive nature, oxygen is not typically stable in terrestrial planet atmospheres and tends to quickly react at the surface by stripping electrons from \emph{reducing} substances in a process known as \emph{oxidation}. Oxygen on Earth is continuously produced by photosynthesis; in the absence of a steady source, continental weathering, volcanic outgassing of reducing gases, and oxidation of basalt via hydrothermal processes at oceanic ridges would quickly remove most of the atmospheric O$_2$ \citep{LR99}. Recently, \cite{CAT14} compiled a table of all major oxidation processes on Earth, obtaining an estimated total O$_2$ removal rate of $2.21\times 10^{13}$ mol/year, or $\sim 150$ bar/Gyr. Over two-thirds of this removal is due to weathering of surface rocks in a process that continuously oxidizes the Earth's crust.

However, the absorption of hundreds to thousands of bars of O$_2$ could eventually lead to the irreversible oxidation of the surface; processes such as plate tectonics, volcanic resurfacing, or volcanic plumes are therefore necessary to subduct the oxidized species and/or supply fresh reductants to the surface. \cite{CAT01} show that the modern rate of (oxidized) Fe$^{3+}$ subduction is equivalent to a removal rate of $5.0\times 10^{11}-1.9\times 10^{12}$ mol O$_2$/year, or $3-12$ bars/Gyr; volcanic outgassing contributes an extra $\sim$ 15 bars/Gyr \citep{CAT14}.

Thus, while a reducing surface can efficiently remove tens and possibly hundreds of bars of O$_2$ on Gyr timescales, the removal will eventually be bottlenecked by the rate of surface recycling. Given that neither Mars nor Venus possess an active tectonic cycle, it is possible that many exoplanets lack plate tectonics, which could significantly delay the removal of O$_2$ from their atmospheres. 

\subsubsection{Water Worlds}
The large weathering rates discussed above assume continental coverage similar to Earth's. Provided they are not completely desiccated, planets that form with tens to hundreds of TO could have significantly more surface water and less continental coverage than Earth after the runaway phase, potentially allowing for longer-lived O$_2$ atmospheres.

But could the oceans act as an oxygen sink? Oxygen is soluble in seawater, saturating at about 8 ml/L ($3.8\times 10^{-4}$ mol/L) at 0$^\circ$ C and 35\permil\ salinity \citep{LEV82}. According to Henry's law, the solubility is proportional to the partial pressure of the gas in equilibrium with the liquid; given a partial pressure of 0.21 bar of O$_2$ on Earth, this corresponds to roughly 0.015 bar of dissolved O$_2$ per bar of atmospheric O$_2$ (for a planet with 1 TO of water). Scaling this to different ocean masses $m_\mathrm{ocean}$, we have
\begin{align}
\frac{\mathrm{mass\ of\ dissolved\ O_2}}{\mathrm{mass\ of\ O_2\ in\ atmosphere}} = 0.015\times\left(\frac{m_\mathrm{ocean}}{1\ \mathrm{TO}}\right),
\end{align}
implying that a terrestrial planet would need roughly 70 TO to absorb half of its atmospheric oxygen into the oceans.

Oxidation of rocks at the seafloor could further deplete the atmospheric O$_2$, but this would require efficient mixing of the oxygen to great depths, which may be difficult for planets with deep oceans. Moreover, tectonic activity or volcanic resurfacing may still be necessary to subduct the oxidized rocks and sustain a long-term surface reductant flux. We thus note that, in general, water worlds may take longer to remove a given amount of O$_2$ from their atmospheres.

\afterpage{
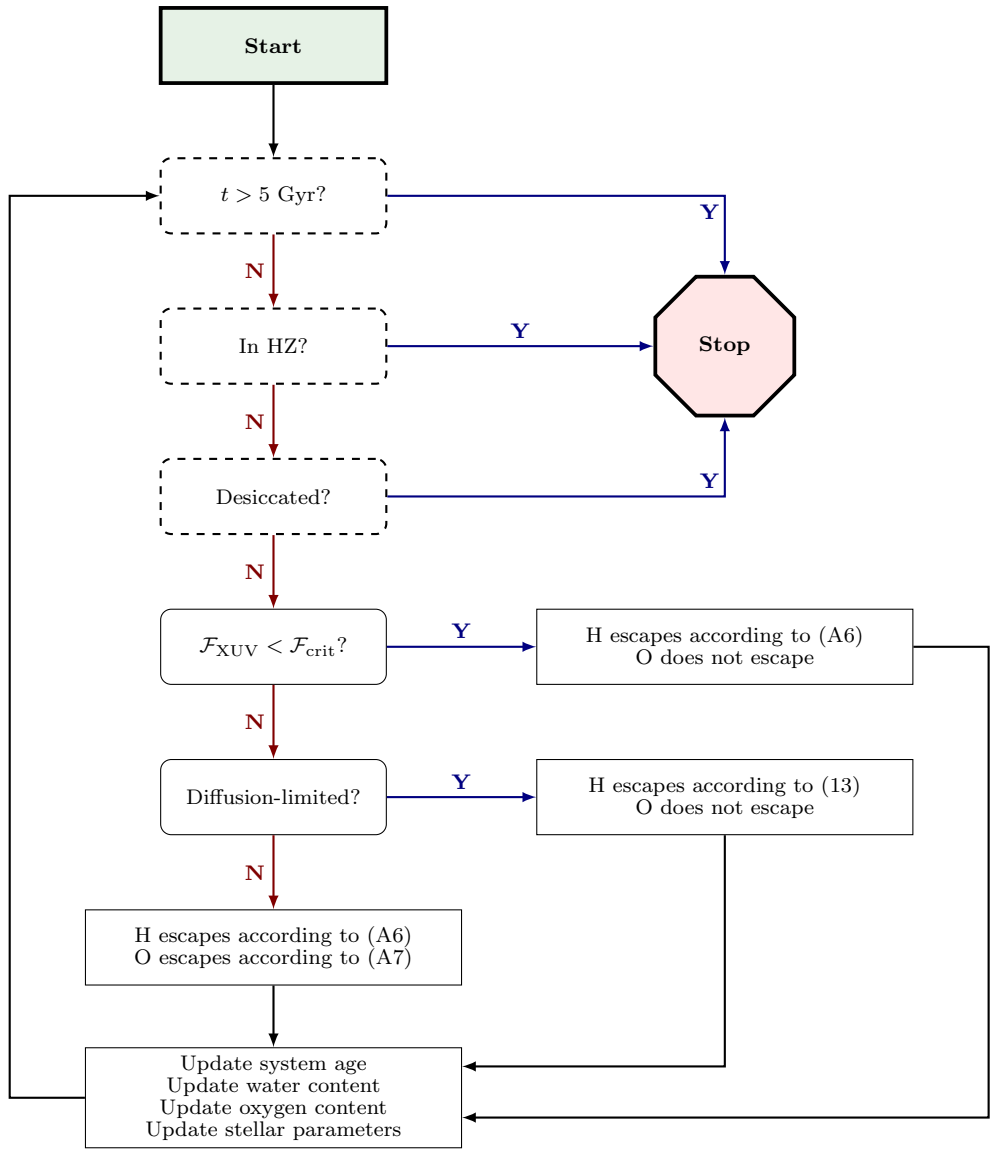
\begin{figure*}[t]
	\begin{center}
	\vspace*{0.4in}
	\tikzstyle{start} = [rectangle, minimum width=3cm, minimum height=1cm,text centered, draw=black, fill=green!50!black!10!white, line width=0.5mm]
	\tikzstyle{stop} = [regular polygon, regular polygon sides=8, minimum width=2cm, minimum height=1cm,text centered, draw=black, fill=red!10!white, line width=0.5mm]
	\tikzstyle{action} = [rectangle, minimum width=5cm, minimum height=1cm, align=center, draw=black, fill=white]
	\tikzstyle{question} = [rectangle, rounded corners, minimum width=3cm, minimum height=1cm, text centered, draw=black, fill=white, dashed, thick]
	\tikzstyle{question2} = [rectangle, rounded corners, minimum width=3cm, minimum height=1cm, text centered, draw=black, fill=white]
	\tikzstyle{arrow} = [thick,->,>=latex]
	\begin{tikzpicture}[node distance=2cm]
		\node (start) [start] {\textbf{Start}};
		\node (age) [question, below of=start] {$t > 5$ Gyr?};
		\node (inhz) [question, below of=age, yshift=0cm] {In HZ?};
		\node (desic) [question, below of=inhz, yshift=0cm] {Desiccated?};
		\node (stop) [stop, right of=inhz, xshift=4cm] {\textbf{Stop}};
		\node (diffusion) [question2, below of=desic, yshift=0cm] {$\mathcal{F}_\mathrm{XUV} < \mathcal{F}_\mathrm{crit}$?};
		\node (difflim) [action, right of=diffusion, xshift=4cm] {H escapes according to (\ref{eq:mdothup})\\O does not escape};
		\node (fcrit) [question2, below of=diffusion, yshift=0cm] {Diffusion-limited?};
		\node (lowfx) [action, right of=fcrit, xshift=4cm] {H escapes according to (\ref{eq:difflim})\\O does not escape};
		\node (highfx) [action, below of=fcrit, yshift=0cm] {H escapes according to (\ref{eq:mdothup})\\O escapes according to (\ref{eq:mdotoup})};
		\node (update) [action, below of=highfx, yshift=0cm] 
			{Update system age\\
			 Update water content\\
			 Update oxygen content\\
			 Update stellar parameters
			};
		\draw [arrow] (start) -- (age);
		\draw [arrow,red!50!black] (age) -- node[anchor=east] {\textbf{N}} (inhz);
		\draw [arrow,blue!50!black] (age) -| node[anchor=north] {\textbf{Y}\ \ \ \ \ } (stop);
		\draw [arrow,red!50!black] (inhz) -- node[anchor=east] {\textbf{N}} (desic);
		\draw [arrow,blue!50!black] (inhz) -- node[anchor=south] {\textbf{Y}} (stop);
		\draw [arrow,blue!50!black] (desic) -| node[anchor=south] {\textbf{Y}\ \ \ \ \ } (stop);
		\draw [arrow,red!50!black] (desic) -- node[anchor=east] {\textbf{N}} (diffusion);
		\draw [arrow,blue!50!black] (diffusion) -- node[anchor=south] {\textbf{Y}} (difflim);
		\draw [arrow,red!50!black] (diffusion) -- node[anchor=east] {\textbf{N}} (fcrit);
		\draw [arrow,blue!50!black] (fcrit) -- node[anchor=south] {\textbf{Y}} (lowfx);
		\draw [arrow,red!50!black] (fcrit) -- node[anchor=east] {\textbf{N}} (highfx);
		\draw [arrow] (highfx) -- (update);
		\draw [arrow] (lowfx) |- ([yshift=1cm]update);	
		\draw [arrow] (difflim) -| ([xshift=1cm]lowfx.east) |- ([yshift=-1cm]update);
		\draw [arrow] (update) -| ([xshift=-1cm]highfx.west) |- (age);
	\end{tikzpicture}
	\caption{Flowchart representing a single integration of our code. The three halting conditions are represented by dashed boxes. See \S\ref{sec:model} for a detailed description of our model.\vspace*{0.1in}}
	\label{fig:flowchart}
	\end{center}
\end{figure*}
}

\subsubsection{Planets With Molten Surfaces}
\label{sec:magmaocean}
We pointed out in \S\ref{sec:planetformation} that because of the high luminosities of M dwarfs early on, planets in the HZ could have magma oceans for extended periods of time following their formation. While this process should not directly affect water loss to space---since several tens of bars of water vapor remain in the atmosphere during the magma ocean phase \citep{MA86,ZAH88,HAM13}---it may prevent the accumulation of atmospheric oxygen. Based on measurements of oxygen diffusion in magma by \cite{WEN91}, \cite{GIL09} argue that photolytically produced oxygen on Venus could diffuse to a depth of $\sim 1$ km over 100 Myr---certainly not enough to absorb all of Venus' oxygen, as this would require an oxidation depth of hundreds of km. However, a convecting magma ocean with a vertical mixing scale of order the mantle thickness ($\sim 3000$ km) could effectively remove all of the atmospheric oxygen (on the order of several hundred bars) on Venus during its early runaway period.

However, the mantle solidification process on planets around M dwarfs is probably different from that on Venus, given the steady decrease in the stellar luminosity with time. While early on these planets receive stellar fluxes several times that received by Earth, during the later stages of the runaway the lower stellar flux could lead to the solidification of most of their mantles, potentially allowing for the buildup of O$_2$ in the atmosphere.

Nevertheless, it is reasonable to expect that planets with Earth-like compositions may be able to absorb a large fraction of the photolytically produced O$_2$ into a primitive magma ocean, at least in the early stages of the runaway, provided (i) the surface composition is similar to Earth's, melting at or below $\sim$ 1500 K; (ii) the surface/interior is initially reducing and capable of absorbing large amounts of oxygen; (iii) the magma ocean is deep and convective, with a sufficiently short turnover time. Strong tidal heating could also potentially extend the magma ocean phase and drive rapid resurfacing, which could lead to efficient oxygen removal from the atmosphere.

\subsubsection{The Case of Venus}
\label{sec:venus}
Finally, we consider the specific case of Venus, which may have lost one or more TO of water during its early runaway period \citep{KAS84,KAS88,CHA96,CHA96I,KUL06,GIL09}. This process should have led to the production of several hundred bars of O$_2$, a large fraction of which may have been deposited in the atmosphere \citep[e.g.,][]{GIL09}. Given the negligible oxygen content of the Venusian atmosphere today \citep{CHA97}, one or more effective oxygen sinks must have existed in the past.

While studies disagree on the process responsible for the removal of this oxygen, many plausible mechanisms exist. \cite{CHA97} showed that a strong primitive solar wind could have heated the upper layers of the Venusian atmosphere, enhancing the thermal escape of hydrogen and facilitating the hydrodynamic drag on the oxygen, potentially carrying all of it to space. \cite{KUL06}, on the other hand, argue that nonthermal interactions between O$^+$ ions and the solar wind could have removed 1 TO of oxygen ($\sim$ 240 bars) over 4.6 Gyr. Surface processes may have also contributed, but \cite{RC95} show that tectonic activity $\sim 15$ times more vigorous than on Earth would be required to subduct all the atmospheric oxygen. \cite{GIL09} argue that vigorous outgassing of reduced gases could remove the equivalent of 1 TO of O$_2$ in 4 Gyr. Finally, a magma ocean could have removed most or all of the O$_2$ \citep{HAM13}, though rigorous quantitative studies of this process are lacking.

We thus emphasize that the removal of a few hundred bars of O$_2$ during/after the runaway greenhouse period is entirely plausible, but highly dependent on properties of the star-planet system. In a review of the evolution of Venus, \cite{CHA12} argue that for different initial conditions, Venus could have had an O$_2$-rich atmosphere for $\sim$ 1 Gyr, and that such planets may exist around other stars.

\section{Model Description}
\label{sec:model}

\begin{figure*}[t]
	\centering   
	\subfigure{
		\label{fig:hzevol3}
		\includegraphics[width=3.25in]{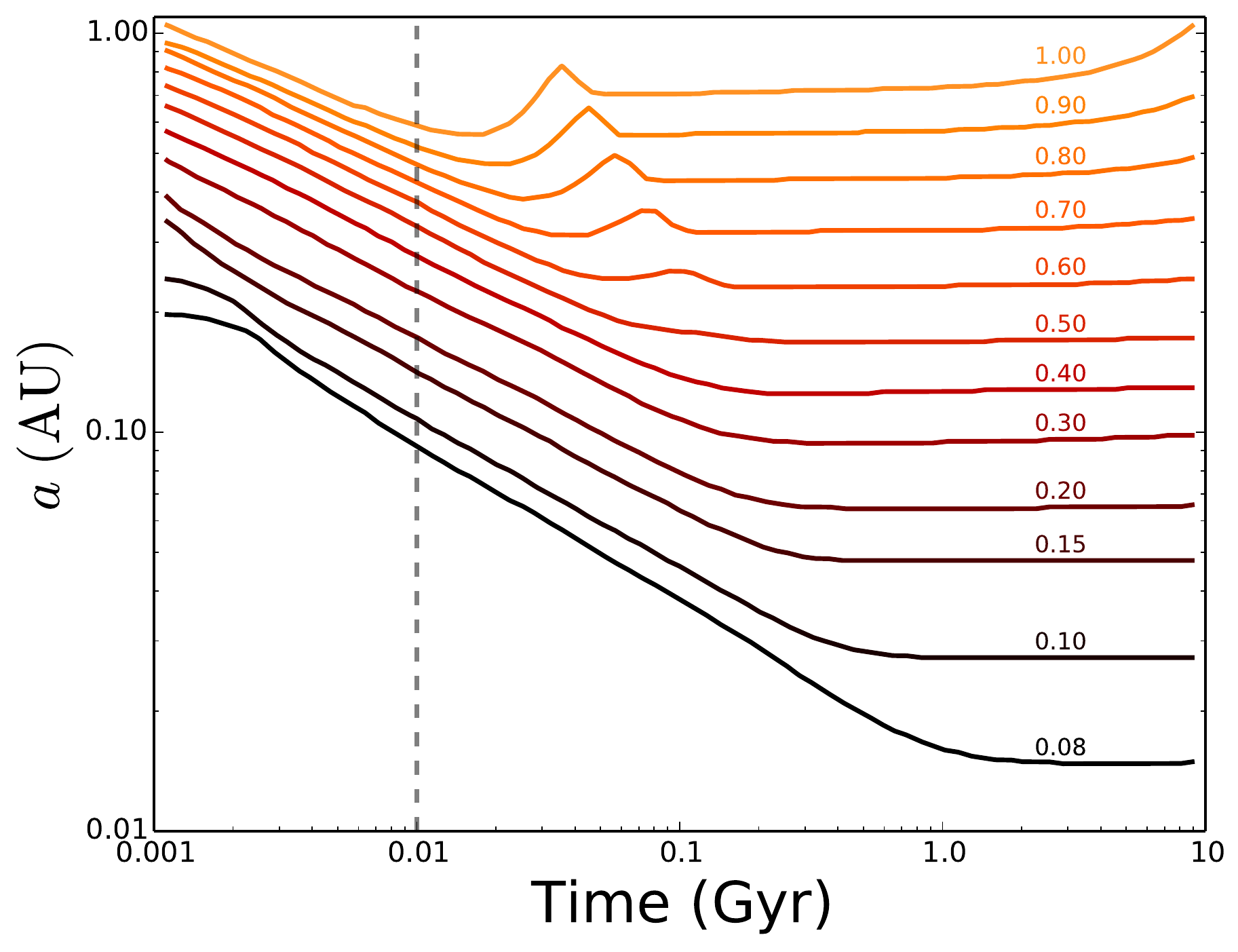}
	}
	\subfigure{
		\label{fig:hzevol2}
		\includegraphics[width=3.25in]{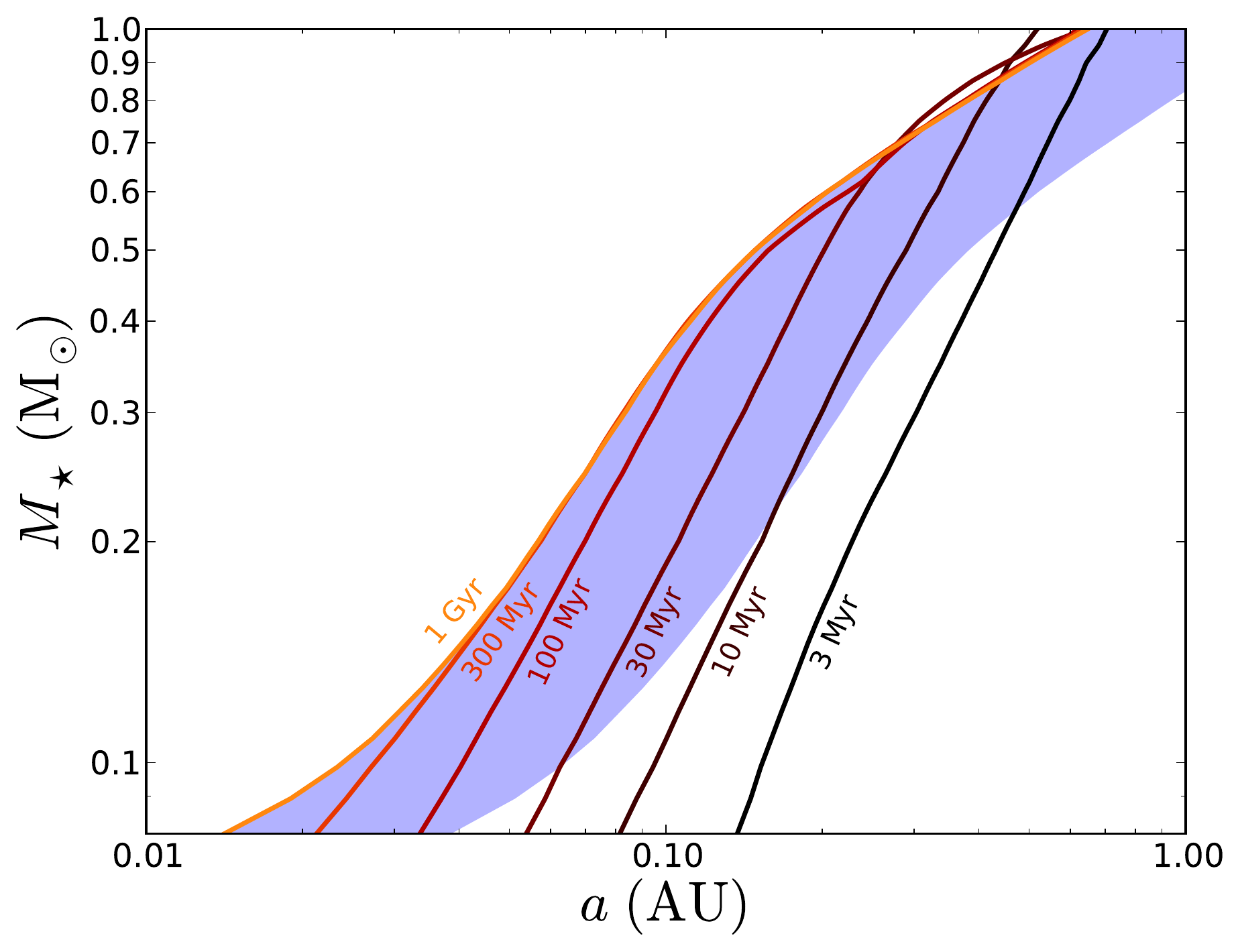}
	}
	\caption{Evolution of the position of the inner edge of the empirical HZ (RV limit) as a function of time. \textbf{Left:} Location of the RV limit versus stellar age for stars between 0.08 M$_\odot$ and 1 M$_\odot$. The vertical dashed line corresponds to the 10 Myr formation time assumed in our model. \textbf{Right:} Contours of the RV limit on a classical HZ plot between 3 Myr (black line) and 1 Gyr (orange line). The empirical HZ at 1 Gyr is shaded in blue for reference. Note that the inner edge moves in by about an order of magnitude for the lowest mass M dwarfs.\vspace*{0.1in}
	\label{fig:hzevol}}
\end{figure*}

For varying planet masses, initial water content, formation times, and HZ limits, we evolve planet/star systems forward in time for 5 Gyr on a grid of stellar mass and semi-major axis, keeping track of the duration of the runaway greenhouse phase, the amount of water lost and the total O$_2$ buildup. As we outline below, we run separate cases assuming either energy-limited escape or diffusion-limited escape. Integrations are performed using an adaptive time-stepping scheme similar to that in \cite{BAR13}.

We consider planets with masses 1 and 5 $\mathrm{M_\oplus}$, with radii calculated from \cite{FMB07} for an Earth-like planet composition (2/3 silicate rock, 1/3 iron). We assume these planets have no significant H/He envelopes and that background atmospheric gases are negligible. The latter is justified by the fact that during a runaway greenhouse on a planet with 1 TO, the atmosphere contains about 270 bars of water vapor \citep{KAS88}; a background atmosphere similar to that on the present Earth is negligible. We further assume zero eccentricity and no tidal evolution for the planets in our runs. We discuss how these assumptions could change our results in \S\ref{sec:discussion}.

We assume a formation time of 10 Myr for all planets. As we discuss in \S\ref{sec:planetformation}, this is probably an upper limit to the migration timescale for planets that form beyond the snow line. As for planets that form \emph{in situ} in the HZ, 10 Myr is roughly equal to the formation time around a $\sim 0.3\mathrm{M}_\odot$ star in the simulations of \cite{RAY07}, but is significantly longer than that predicted by the scaling arguments in \cite{LIS07}. Our adopted value is therefore likely to overestimate the formation time for planets around low mass M dwarfs. Around the highest mass M dwarfs and K dwarfs ($\gtrsim 0.5\mathrm{M}_\odot$), we are likely underestimating the formation time. In both formation scenarios, we vary the initial surface water content between 1 and 100 TO, noting that planets that migrate from beyond the snow line are more likely to have $\gtrsim 10$ TO.

At each timestep, we calculate the HZ limits for 1M$_\oplus$ and 5M$_\oplus$ planets from \cite{KOP14}, using $L_\mathrm{bol}$ and $T_\mathrm{eff}$ from the cooling tracks of \cite{BAR98} for solar metallicity. We calculate water loss and O$_2$ buildup as long as planets are in a runaway greenhouse. Given the variety of mechanisms capable of removing atmospheric O$_2$ and their complex dependence on an array of planetary properties (\S\ref{sec:o2atmospheres}), we do not model the details of oxygen absorption by the surface. Instead, we consider two limiting cases regarding the oxygen: \textbf{(a)} efficient absorption by the surface, corresponding to an effectively instantaneous removal of O$_2$ by surface processes; and \textbf{(b)} inefficient absorption, corresponding to a rate of oxygen buildup much larger than the rate at which it is removed. 

In case (a), which could be the case of a planet with a deep magma ocean, we assume the atmosphere is predominantly H$_2$O and that diffusion through an oxygen-rich layer does not take place; we therefore use the energy-limited escape equations (\ref{eq:mdothup})-(\ref{eq:mdotocean}) to calculate H and O loss rates and O$_2$ buildup rates. In case (b), corresponding to (say) a planet with a highly oxidized surface and/or one that lacks plate tectonics, we assume that all of the O$_2$ remains in the atmosphere. This should quickly result in a depletion of H and H$_2$O relative to oxygen in the upper layers of the atmosphere, such that hydrogen escape will be limited by diffusion. We therefore set the hydrogen escape flux to the \emph{minimum} of the energy-limited escape flux (\ref{eq:fhfhref}) and the diffusion-limit escape flux (\ref{eq:difflim}). In (\ref{eq:fhfhref}), the ratio $X_\mathrm{O}/X_\mathrm{H}$ is calculated from the ratio of H$_2$O to O$_2$ in the atmosphere, assuming the two species are well-mixed below the diffusion layer. In this regime, the oxygen escape flux is assumed to be zero.

We report the amount of oxygen retained by the planet (either in the atmosphere or absorbed by the surface) as an equivalent pressure in bars, which we define to be the surface pressure of the oxygen \emph{if it were the only gas in the atmosphere}. Thus, an equivalent pressure of 1 bar corresponds to an amount of oxygen equal in mass to the atmosphere of the current Earth, or about five times the current mass of oxygen in the atmosphere.\footnote{Since the molecular weights of N$_2$ and O$_2$ are similar, the equivalent pressure of O$_2$ on the present-day Earth is very close to its partial pressure.} For a given planet, the \emph{partial} pressure corresponding to this amount will depend on the mixing ratios and the mean molecular weights of other species in the atmosphere. Therefore, to preserve generality, below we present the equivalent O$_2$ pressure in all of our figures.

In order to capture the uncertainty in the critical stellar flux above which planets go runaway, we also run two separate sets of models: one in which the runaway occurs interior to the RG limit (the default case), and one interior to the RV limit. Since terrestrial planets with surface oceans are likely to enter the runaway phase somewhere in between these limits, the two runs should roughly bracket the actual evolution.

For simplicity, we use a saturation time of 0.1 Gyr for K dwarfs ($M_\star > 0.6\mathrm{M_\odot}$) and 1 Gyr for M dwarfs. We assume an XUV absorption efficiency in the range $0.15 \leq \epsilon_\mathrm{XUV} \leq 0.30$, typical of hydrogen-rich atmospheres \citep{CHA96,WP13}. It is likely that the efficiency for an H$_2$/H$_2$O atmosphere is on the high end of this interval, since the range of wavelengths capable of heating the gas is larger due to absorption by H$_2$O in the FUV \citep{SEK81}; unless otherwise noted, we take $\epsilon_\mathrm{XUV} = 0.30$. As in \cite{WP14}, we further assume that water loss is limited by the escape of H to space rather than by the H$_2$O photolysis rate.

We note, finally, that our procedure may lead to an underestimate of the total water loss. Even after a planet leaves the runaway greenhouse, diffusion of water vapor into the stratosphere and H escape can still be significant; for instance, \cite{WP14} show that about 28\% of a TO can be lost from an N$_2$-poor Earth in the HZ over 4 Gyr.

A flowchart illustrating a sample integration of our code is presented in Figure~\ref{fig:flowchart}. The integration continues as long as all three halting conditions (represented by the dashed boxes) evaluate to false. The code halts if \textbf{(1)} the age of the system reaches 5 Gyr; \textbf{(2)} the planet enters the HZ (corresponding to the end of the runaway greenhouse phase); or \textbf{(3)} the planet is completely desiccated. Otherwise, the code calculates H and O loss rates and O$_2$ buildup rates as outlined above. The code then updates stellar and planetary parameters and loops.

\section{Results: Evolution of the Habitable Zone}
\label{sec:results1}
\subsection{Location of the Habitable Zone}
In Figure~\ref{fig:hzevol} we show the evolution of the HZ as a function of time for M, K, and G dwarfs. In the left panel, we plot the position of the RV limit as a function of stellar age for stars ranging in mass from 0.08 M$_\odot$ to 1 M$_\odot$. For solar-type G stars, the HZ evolves inwards primarily in the first 10 Myr, prior to the formation of terrestrial planets. For K dwarfs ($0.6\mathrm{M}_\odot \lesssim M_\star  \lesssim 0.9\mathrm{M}_\odot$), only planets that form in the first $\sim$ 10 Myr experience a significant change in the location of the HZ. The HZ of M dwarfs, on the other hand, moves in significantly, even after the formation of terrestrial planets. Around the lowest mass M dwarfs, the inner edge of the HZ moves in by nearly an order of magnitude after 10 Myr.

In the right panel, we plot the RV limit at different stellar ages on a classical HZ plot; the extent of the HZ at 1 Gyr is indicated by the blue shading. Note that even planets located at the outer edge of the HZ at $t > 1$ Gyr were interior to the HZ early on, especially for low mass M dwarfs. The change in the location of the HZ of solar-type stars is comparatively small. For stars more massive than about 0.8 M$_\odot$, the HZ moves inward during the PMS phase and then outward due to the steady increase of $L_\mathrm{bol}$ of these stars once they are on the MS. The HZ of M dwarfs, on the other hand, moves monotonically inward for up to 1 Gyr. After this point, the HZ boundaries remain relatively stationary for up to tens of Gyr due to these stars' slow MS evolution.

\begin{figure}[h!]
	\centering  
	\label{fig:gtime100}
	\includegraphics[width=3.5in]{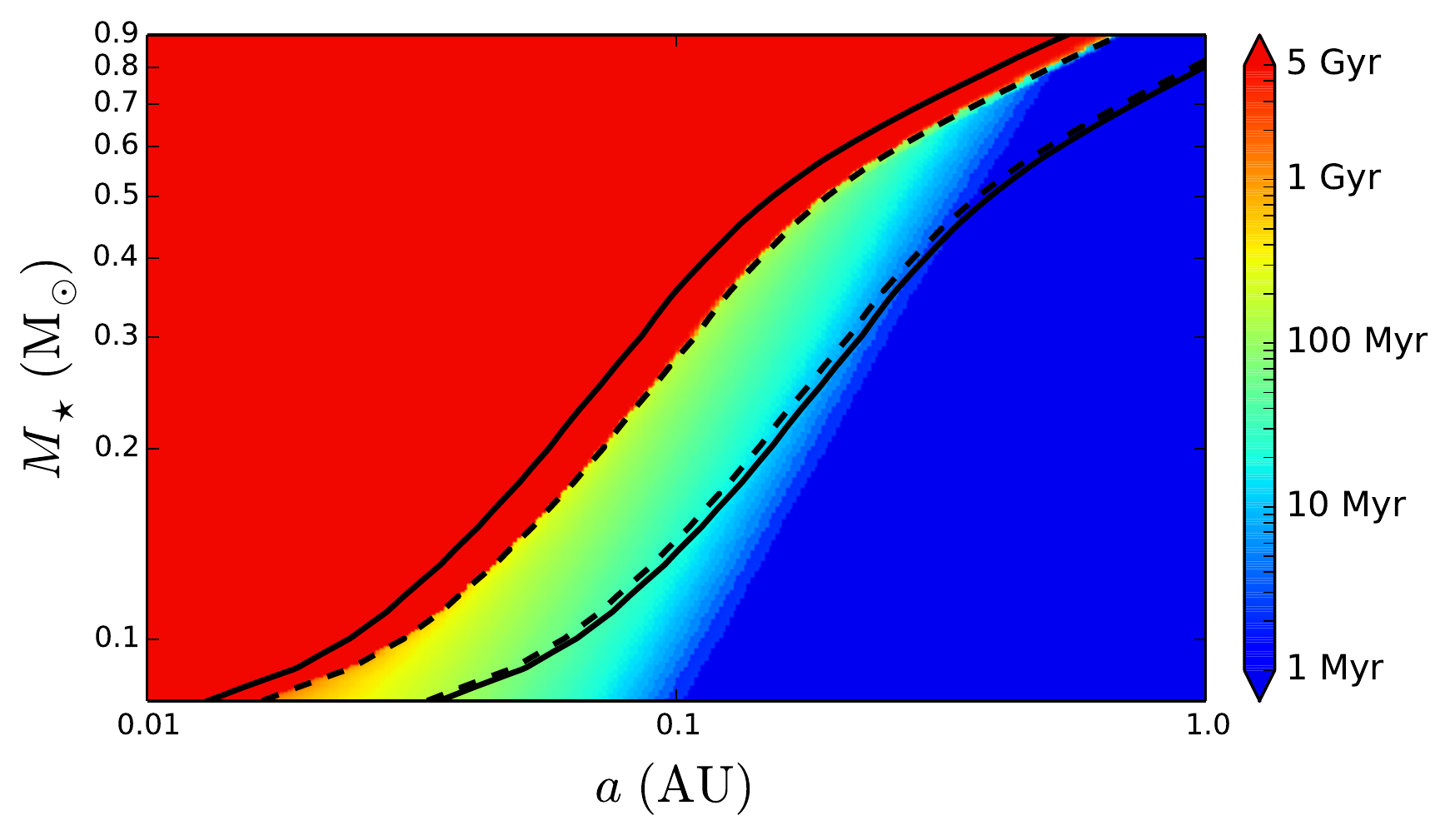}
	\caption{Duration of the runaway greenhouse for planets that formed at 10 Myr with abundant surface water. The solid black lines correspond to the RV (left) and EM (right) limits (the empirical HZ); the dashed lines correspond to the RG (left) and the MG (right) limits (the theoretical HZ). All limits are those at a stellar age of 5 Gyr.  Red corresponds to planets that never leave the runaway state, since these are always interior to the HZ. Blue corresponds to planets that spend less than 1 Myr in a runaway greenhouse. Planets throughout the HZs of all M dwarfs spend a substantial amount of time in a runaway greenhouse.\vspace*{0.1in}
	\label{fig:gtime}}
\end{figure}

\subsection{Duration of the Runaway Phase}
Figure~\ref{fig:gtime} shows the duration of the runaway phase for planets that form at 10 Myr, assuming the runaway occurs interior to the RG limit. The vertical axis (stellar mass) ranges from late M dwarfs (0.08 M$_\odot$) to early K dwarfs (0.9 M$_\odot$). The horizontal axis (semi-major axis) spans the HZ, which is bounded on the left and right by the RV and EM limits (solid lines); from left to right, the RG and MG limits are indicated by the dashed lines. The limits are those at a stellar age of 5 Gyr.

Unsurprisingly, interior to the RG limit, planets spend the entire age of the system (5 Gyr) in a runaway state. Note that once a planet's water is depleted, it is technically no longer in a runaway greenhouse, since the atmospheric infrared windows will open up and the surface will cool. For the purposes of this figure, we therefore assume an unlimited surface water inventory.

Far to the right of the HZ, planets are never in an insolation-induced runaway greenhouse. However, throughout most of the HZ of M dwarfs, planets spend tens to hundreds of Myr in an early runaway greenhouse phase. Planets in the HZs of K dwarfs may experience a runaway greenhouse for a few tens of Myr, but only if they form early. Above about 0.8 M$_\odot$, the duration is negligible, except in the vicinity of the RG boundary; this is the case for planets around solar-type stars.

Because of this early runaway phase, many planets orbiting in the HZs of M dwarfs could potentially remain permanently uninhabitable. In the next section, we investigate the effect that the runaway greenhouse has on water loss and oxygen buildup on these planets.

\section{Results: Water Loss and O$_2$ Buildup}
\label{sec:results2}

\begin{figure*}[t]
	\centering   
	\textbf{\large \sc Energy-Limited Escape: 1 $\mathrm{M}_\oplus$, 1 TO}\par\medskip
	\subfigure[\textbf{Water Lost}]{
		\includegraphics[width=3.25in]{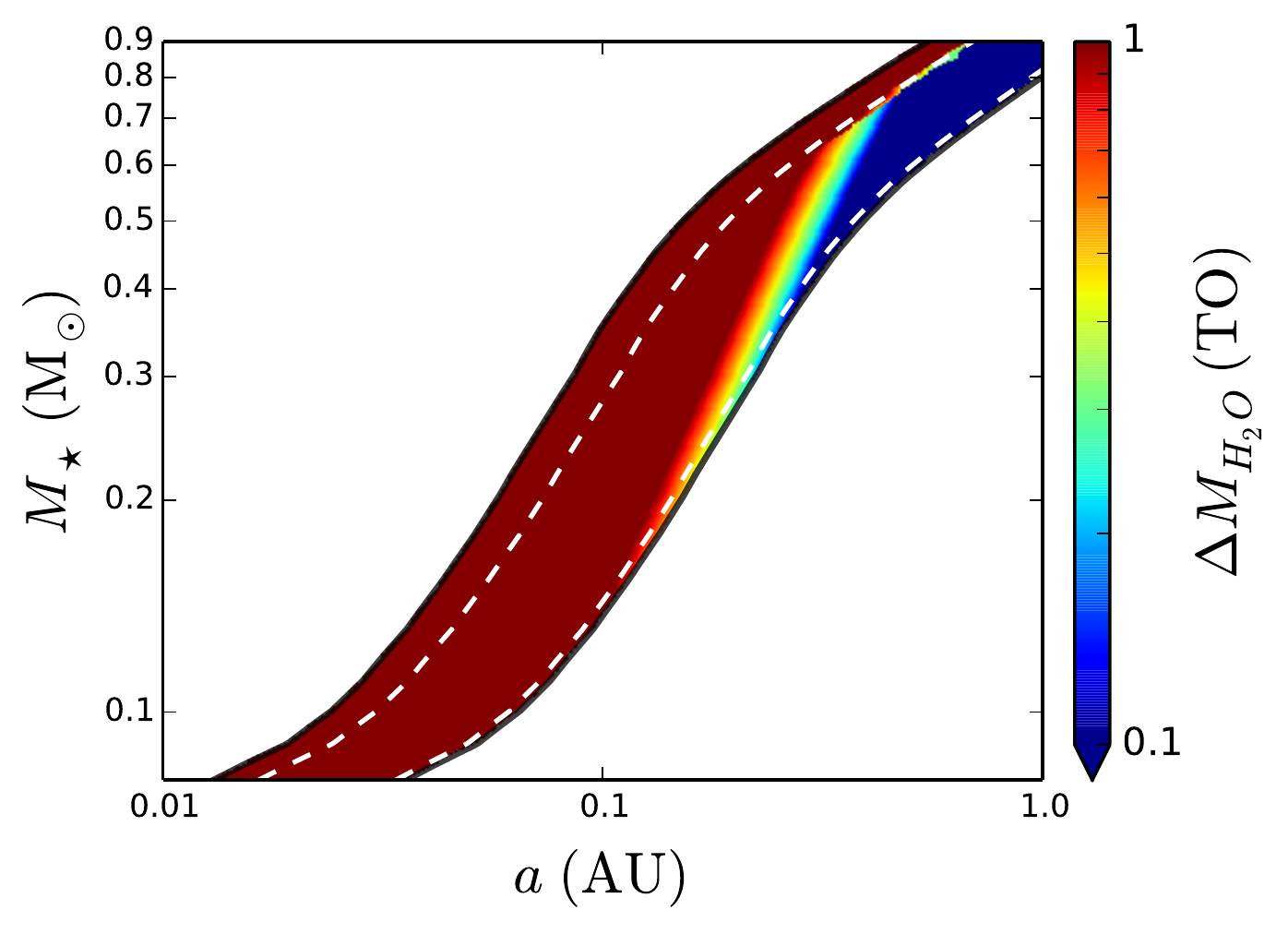}
	}
	\subfigure[\textbf{Oxygen Absorbed by Surface}]{
		\includegraphics[width=3.25in]{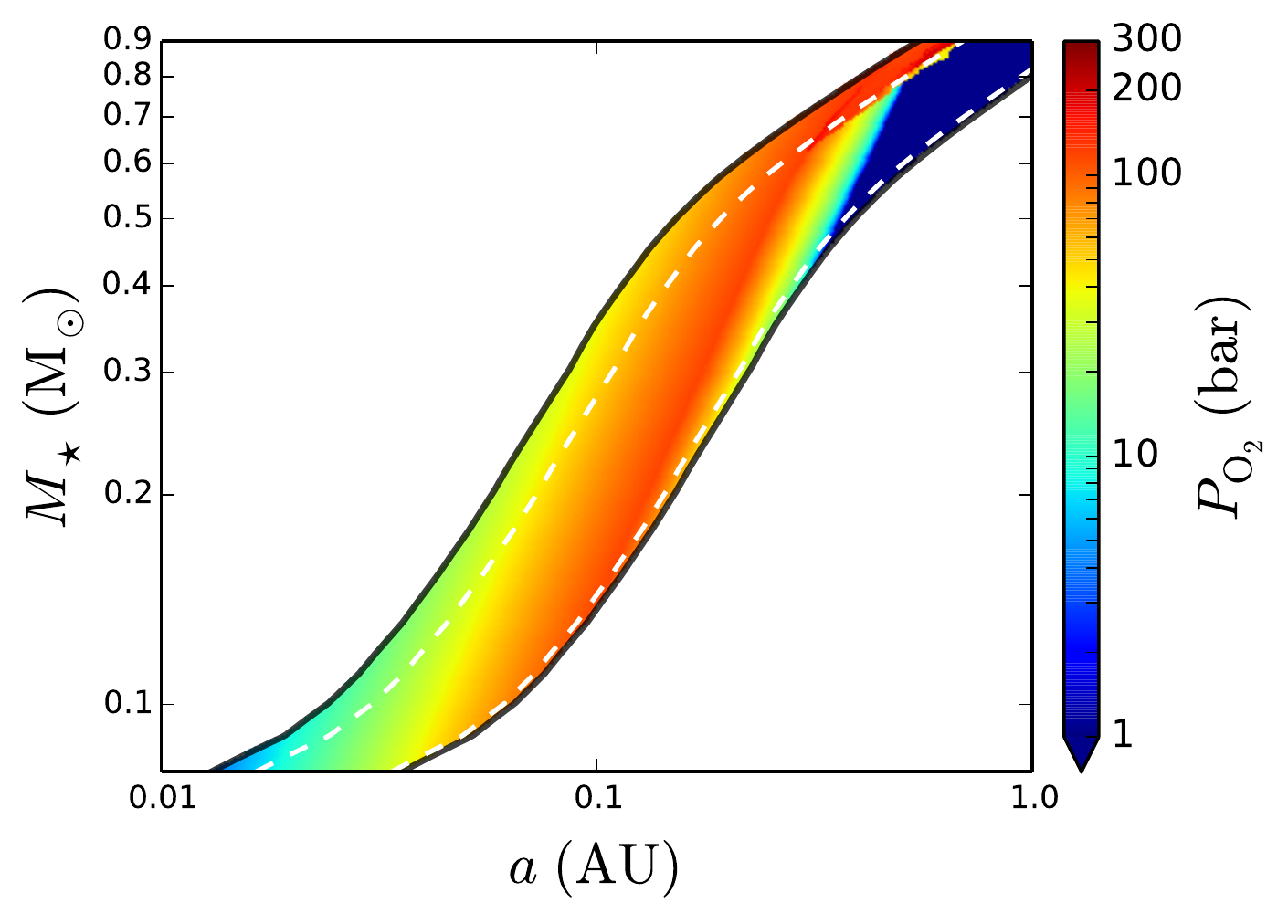}
	}
	\caption{Total amount of water lost and amount of oxygen absorbed at the surface for a 1M$_\oplus$ planet formed at 10 Myr with 1 TO of surface water, assuming the planet is in a runaway interior to the RG limit, the oxygen is instantaneously absorbed by the surface, and the escape is energy-limited. The solid lines are the empirical HZ bounds; the dashed lines are the theoretical HZ bounds. See Figure~\ref{fig:elim1TO} for more details.\vspace*{0.1in}
	\label{fig:hzoverlay}}
\end{figure*}

\begin{figure*}[t]
	\centering 
	\textbf{\large \sc Energy-Limited Escape: 1 $\mathrm{M}_\oplus$, 1 TO}\par\medskip    
	\subfigure[\textbf{Water Lost}]{
		\includegraphics[width=3.25in]{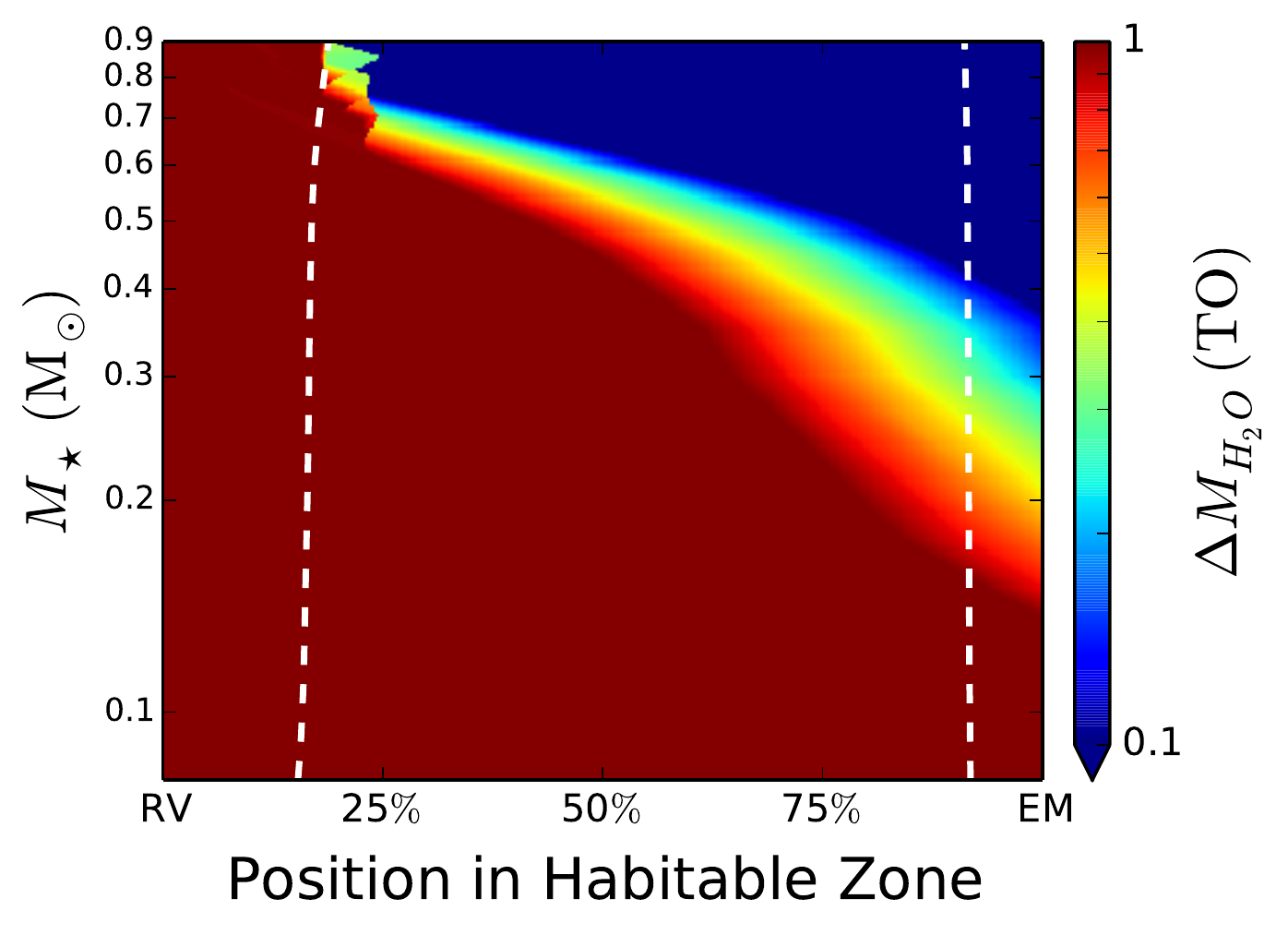}
	}
	\subfigure[\textbf{Oxygen Absorbed by Surface}]{
		\includegraphics[width=3.25in]{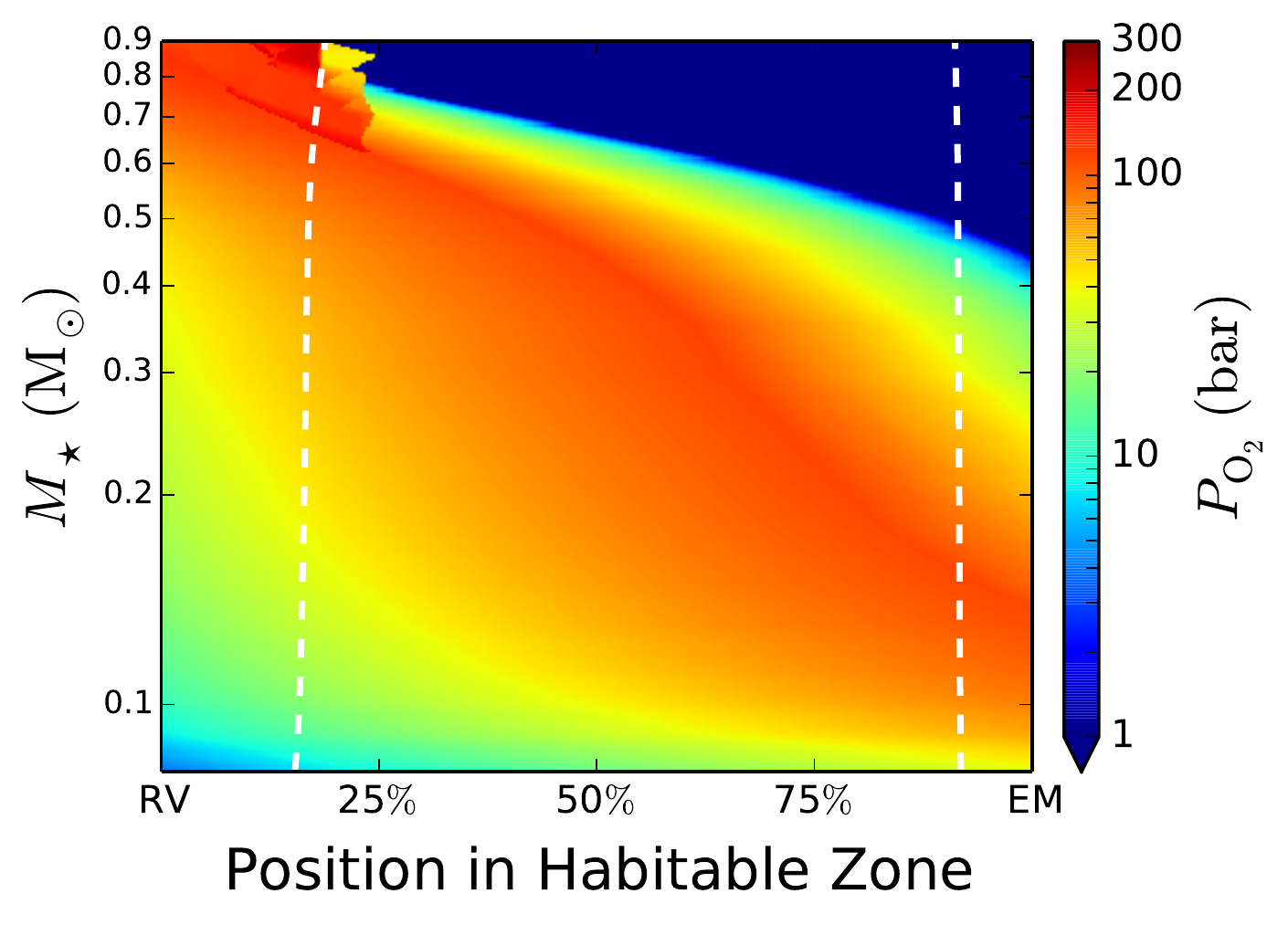}
	}
	\caption{Same as Figure~\ref{fig:hzoverlay} (1 TO, energy-limited), but in an expanded view of the HZ. The axes correspond to the stellar mass (vertical) and the position of the planet within the HZ at 5 Gyr (horizontal). The ``position in habitable zone'' is the fractional distance between the RV limit and the EM limit (the empirical HZ). The dashed lines once again represent the RG and MG limits. \textbf{(a)} Total water lost in TO after 5 Gyr. Dark blue corresponds to less than 0.1 TO; dark red corresponds to complete desiccation. Most planets in the HZ of M dwarfs are completely desiccated; conversely, those close to the outer edge of high mass M dwarfs and throughout most of the HZ of K dwarfs lose little or no water. Interior to the RG limit, planets around stars of all masses are completely desiccated. \textbf{(b)} Total amount of oxygen absorbed by the surface in bars. Dark blue corresponds to insignificant O$_2$ buildup; dark red corresponds to 200 bars of oxygen. Planets that lose significant amounts of water also undergo extreme surface oxidation.\vspace*{0.1in}
	\label{fig:elim1TO}}
\end{figure*}

\begin{figure*}[t]
	\centering  
	\textbf{\large \sc Energy-Limited Escape: 1 $\mathrm{M}_\oplus$, 10 TO}\par\medskip  
	\subfigure[\textbf{Water Lost}]{
		\includegraphics[width=3.25in]{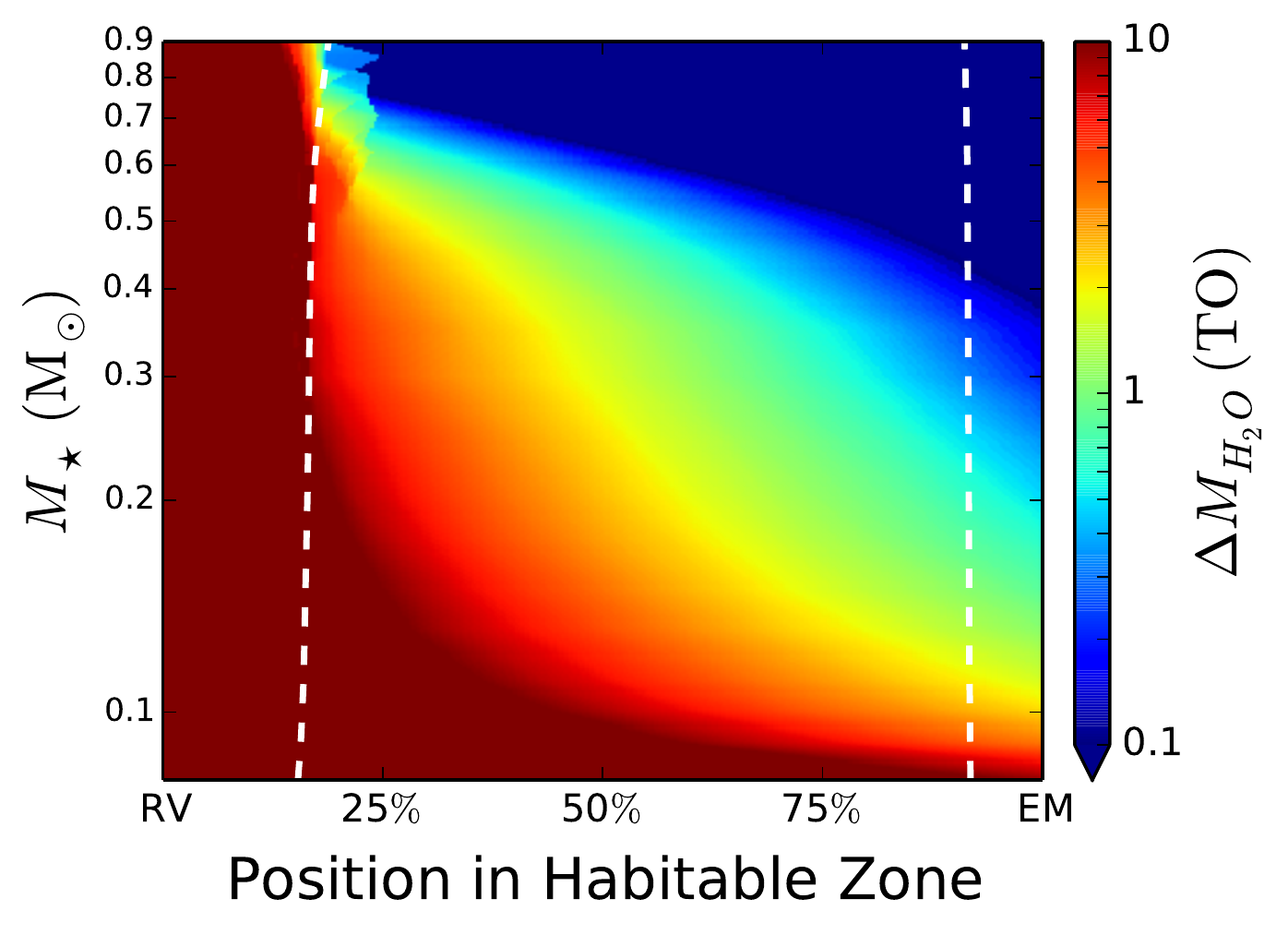}
	}
	\subfigure[\textbf{Oxygen Absorbed by Surface}]{
		\includegraphics[width=3.25in]{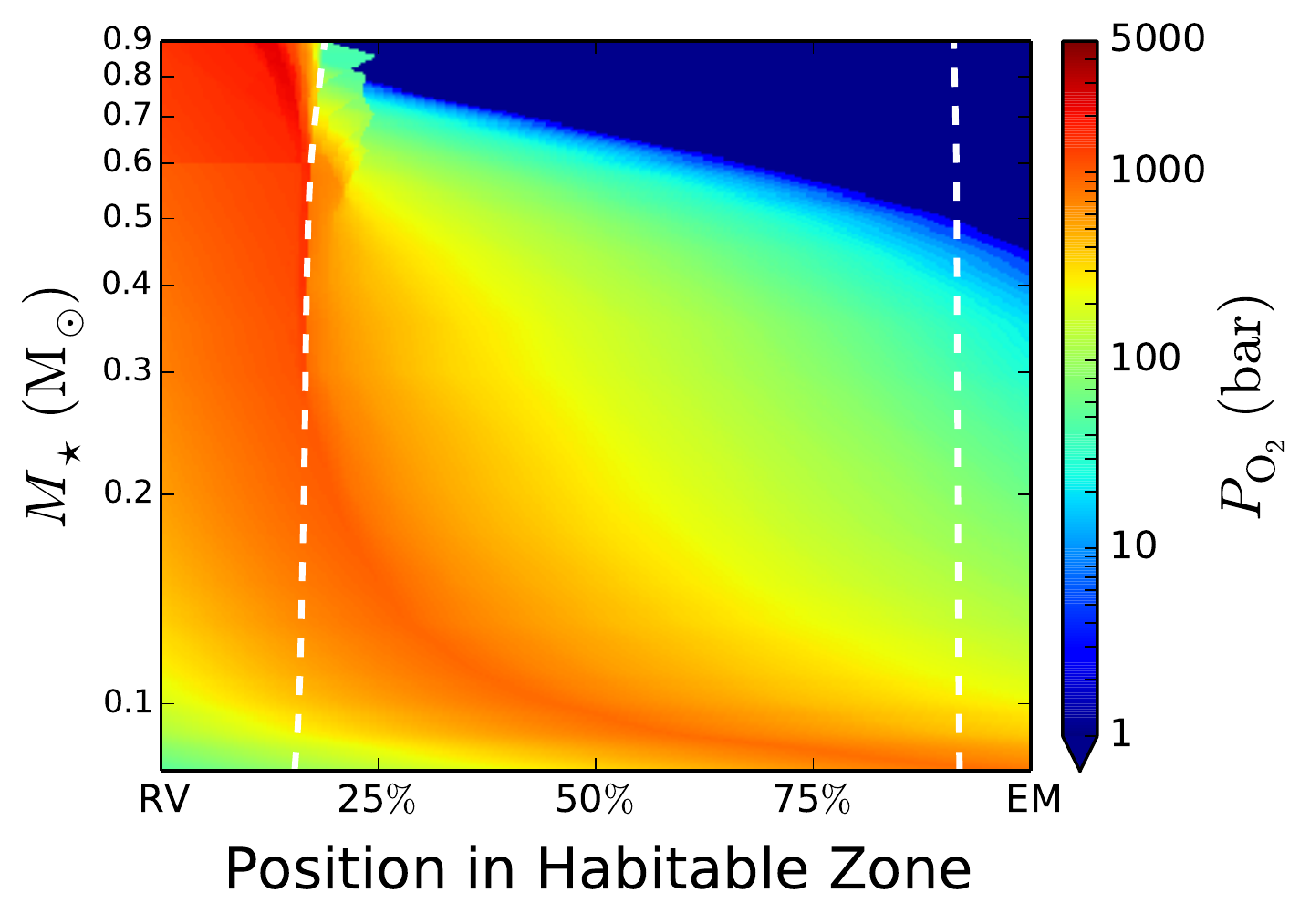}
	}
	\caption{Same as Figure~\ref{fig:elim1TO} (energy-limited escape), but for an initial water content of 10 TO. Note the change in the colorbar scales. Planets throughout most of the HZ of M dwarfs now lose at least 1 TO of water; those close to the RG limit and around low mass M dwarfs lose close to 10 TO. Most planets now retain several hundred to $\sim$ 1000 bars of O$_2$.\vspace*{0.1in}
	\label{fig:elim10TO}}
\end{figure*}

\subsection{Validation Against Venus}
Before we present our results for M dwarf planets, we briefly examine water loss and O$_2$ buildup on Venus. Assuming an XUV saturation fraction $f_0 = 10^{-3}$, a saturation timescale of 0.1 Gyr, a formation time of $50$ Myr, an initial water content of 1 TO \citep{RAY06}, and a runaway greenhouse interior to the RG limit, we find that Venus is completely desiccated in both the energy-limited and diffusion-limited escape regimes, accumulating an equivalent O$_2$ pressure between $\sim$ 120 bars (energy-limited) and $\sim$ 240 bars (diffusion-limited). For a much larger initial water content of 10 TO, it is still completely desiccated in both regimes and builds up between $\sim 1800$ and $\sim 2400$ bars of O$_2$. If, on the other hand, we assume the more optimistic HZ boundary, such that a runaway greenhouse occurs only once Venus is interior to the RV limit, Venus loses a maximum of 0.5 TO and builds up a maximum of 120 bars of O$_2$ in both regimes. 

These figures are broadly consistent with previous estimates \citep{KAS84,KAS88,CHA96,CHA96I,KUL06,GIL09}, though the large uncertainty regarding the initial water content precludes an accurate determination of the amount of O$_2$ retained by the planet. \cite{LAM11a} argue that most of the O$_2$ that does not escape hydrodynamically could be removed by oxygen blow-off, but the exospheric temperatures may not be high enough for this to occur \citep[see, e.g.,][and \S\ref{sec:otherremarks}]{TIA09}. In any event, the negligible O$_2$ content of the Venusian atmosphere today implies that the O$_2$ must have been removed either by surface sinks or other escape mechanisms. Given terrestrial rates of O$_2$ removal of a few hundred bars per Gyr (\S\ref{sec:o2atmospheres}), our values are consistent with a maximum of a few TO of water on early Venus and therefore agree with the studies mentioned above.

\subsection{Fast Oxygen Removal: Energy-Limited Escape}
We first consider the limiting case where the rate of oxygen absorption by surface sinks is much larger than the rate at which it is photolytically produced. This could happen in the case of planets with vigorous resurfacing processes or convecting magma oceans (see \S\ref{sec:magmaocean} and \S\ref{sec:model} for a discussion). The oxygen content of the atmospheres of these planets is always low, the upper atmosphere is rich in water vapor, and hydrogen and oxygen escape at the energy-limited rate (\S\ref{sec:oxygenescape}).

In Figure~\ref{fig:hzoverlay} we plot the results for a 1M$_\oplus$ planet with 1 TO of initial surface water, formed at 10 Myr, assuming a runaway greenhouse occurs interior to the RG limit. The axes are the same as in Figure~\ref{fig:gtime}, but the colors now indicate the total amount of water lost (left panel) and the equivalent pressure of O$_2$ absorbed by the surface in bars (right panel). While Figure~\ref{fig:hzoverlay} shows our results on a traditional HZ plot, we re-scale the $x$-axis to be the relative position in the (empirical) 5 Gyr HZ in Figure~\ref{fig:elim1TO}. This corresponds to zooming in on the HZ at each stellar mass in Figure~\ref{fig:hzoverlay} (note that the data plotted here are exactly the same). For reference, the dashed vertical lines once again indicate the locations of the RG (left) and MG (right) limits.

Planets throughout most of the HZ of M dwarfs are completely desiccated (left panel). Above $\sim 0.2\mathrm{M}_\odot$, planets close to the outer edge of the HZ retain some of their water due to the shorter runaway phase; however, even planets in the center of the HZ of high mass M dwarfs ($M_\star \gtrsim 0.4 \mathrm{M}_\odot$) are completely desiccated. Planets around K dwarfs ($M_\star \gtrsim 0.6 \mathrm{M}_\odot$), on the other hand, lose significant amounts of water only close to the RG limit. Note, importantly, that planets that form \emph{in situ} in the HZs of K dwarfs likely take longer than 10 Myr to form \citep{RAY07}; therefore, above $\sim 0.6 \mathrm{M}_\odot$, this figure applies only to planets that form quickly beyond the snow line and migrate into the HZ. Finally, although the figures encompass only M and K dwarfs, we note that planets in the theoretical HZ of solar-mass stars are never in an insolation-induced runaway greenhouse and therefore lose no water via this mechanism.

Planets that lose significant water also retain on the order of 100 bars of O$_2$ (right panel), which by assumption is quickly absorbed by the surface. Such a high oxidative power could have strong implications for planetary evolution and habitability, which we discuss in \S\ref{sec:discussion}. Perhaps surprisingly, the maximum oxygen pressure actually occurs in the \emph{center} of the HZ of mid- to high mass M dwarfs and close to the \emph{outer} edge for low mass M dwarfs. This is because close to the inner edge of the HZ, the ocean is lost quickly and the planet is desiccated early on, when the XUV flux is high and oxygen escape is efficient. Close to the outer edge, the RG phase is shorter, resulting in weaker ocean loss and similarly low O$_2$ amounts. Therefore, for each stellar mass, there exists a certain distance at which the O$_2$ pressure peaks. We discuss this trend further in \S\ref{sec:constantpodot}.

Next, in Figure~\ref{fig:elim10TO} we repeat the previous calculations for an initial water content of 10 TO. Complete desiccation now occurs only around low mass M dwarfs, particularly close to the RG limit. However, most M dwarf planets lose $> 1$ TO throughout the HZ and $\gtrsim 5$ TO close to the RG limit. Moreover, the equivalent pressure of oxygen absorbed by the surface is now on the order of several hundred to $\sim 1000$ bars, peaking in the outer HZ of the lowest mass M dwarfs and close to the RG limit of $0.3 \mathrm{M}_\odot$ M dwarfs.

\begin{figure}[b]
  \begin{center}
    \leavevmode
      \psfig{file=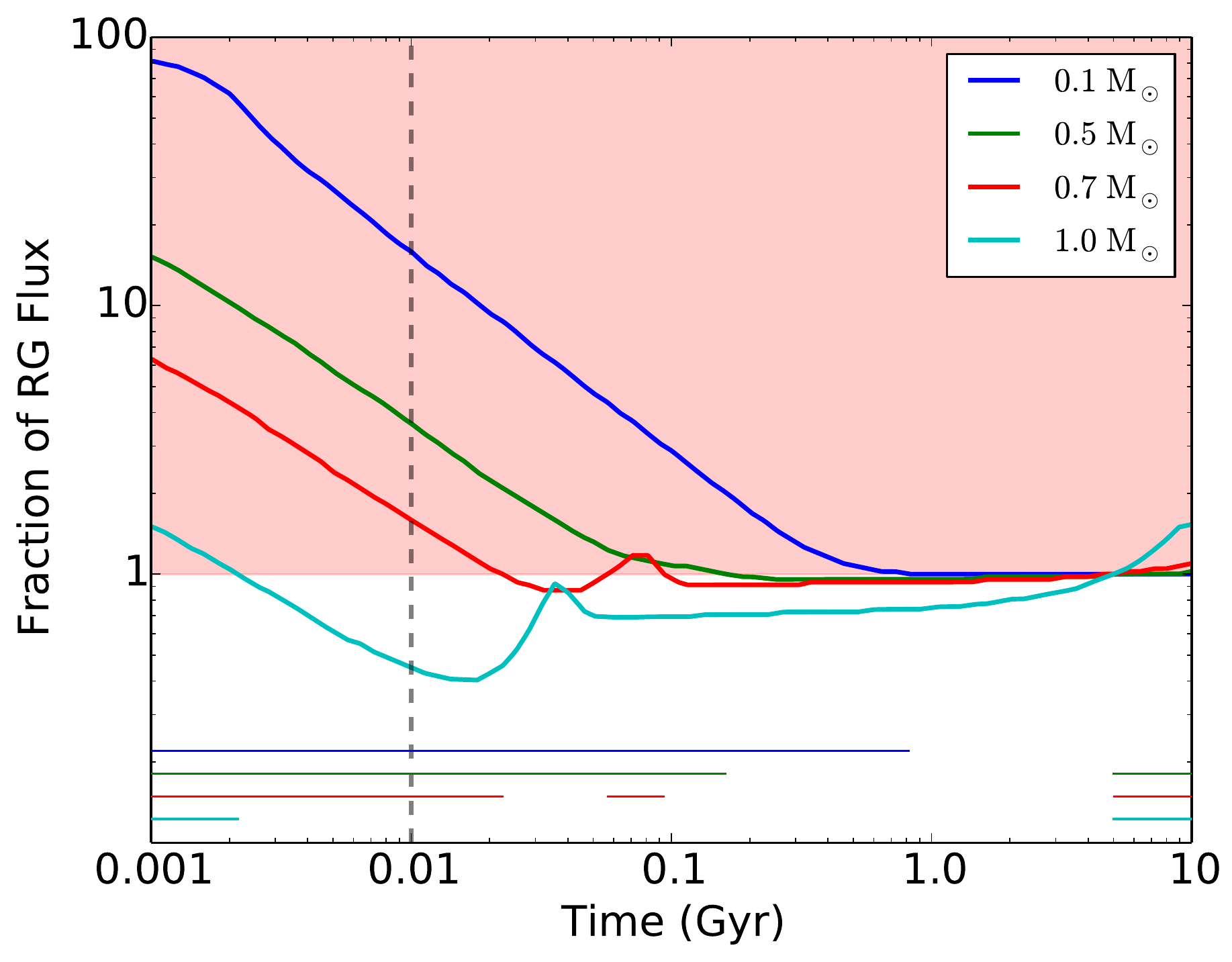,width=3.5in}
       \caption{Evolution of the flux received by planets that are at the inner edge of the theoretical HZ at 5 Gyr for different stellar masses (blue: 0.1M$_\odot$, green: 0.5M$_\odot$, red: 0.7M$_\odot$, cyan: 1.0M$_\odot$). The vertical axis is the bolometric flux normalized to the runaway greenhouse flux; the region shaded in pink corresponds to planets that are interior to the RG limit. The dashed vertical line corresponds to the 10 Myr formation time we assume in our model. The horizontal lines at the bottom of the plot indicate the times during which planets are in a runaway greenhouse. See text for a discussion.
       \vspace*{0.1in}}
     \label{fig:bolfluxevol}
  \end{center}
\end{figure}

Finally, we note that there is an interesting feature near the RG limit of K dwarfs in these figures. Consider Figure~\ref{fig:elim1TO}, for instance: above 0.6M$_\odot$, both the total water loss and the O$_2$ amount change discontinuously as the stellar mass increases, leading to a jagged pattern near the RG limit and high O$_2$ buildup. This behavior is rooted in the non-monotonic luminosity evolution of K dwarfs prior to $\sim$ 100 Myr. In Figure~\ref{fig:bolfluxevol} we plot the evolution of the bolometric flux received by a planet near the inner edge of the HZ (defined at 5 Gyr), in units of the critical runaway greenhouse flux; values above 1 (red shading) correspond to a runaway greenhouse state. The thick curves indicate the flux evolution for different stellar masses (blue: a late M dwarf; green: an early M dwarf; red: a K dwarf; cyan: a solar-type G dwarf). Note that while M dwarfs dim monotonically during their PMS contraction phase, stars of type K and earlier display a bump in their luminosity prior to 100 Myr. This is due to the fact that stars more massive than about 0.6 M$_\odot$ switch from convective to radiative energy transport towards the end of their contraction phase, during which time their effective temperatures rise, leading to a temporary increase in $L$ just before reaching the MS \citep[see, e.g.,][]{RH05}. M dwarfs, which remain mostly convective even once they reach the MS, do not display such a bump.

For a star like the Sun (cyan curve), the bump in the luminosity does not significantly affect planets in the HZ, since the flux on a planet near the inner edge does not exceed the critical flux during that time. For K dwarfs, however, this is not true; the flux on a planet at the inner edge of the HZ around a 0.7 M$_\odot$ star exceeds the critical flux for $\sim 30$ Myr, leading to two distinct runaway greenhouse episodes prior to 5 Gyr.

For reference, the thin lines at the bottom of the plot indicate when planets are in the runaway regime. While M dwarf planets experience the longest runaway greenhouse episodes, planets that form prior to about 20 Myr in the inner HZ of K dwarfs undergo two runaways, the second of which occurs late enough that the XUV flux has largely tapered off, leading to efficient oxygen buildup.

\subsection{Inefficient Oxygen Sinks: Diffusion-Limited Escape}
\label{sec:slowsinks}

\begin{figure*}[t]
	\centering   
	\textbf{\large \sc Diffusion-Limited Escape: 1 $\mathrm{M}_\oplus$, 1 TO}\par\medskip 
	\subfigure[\textbf{Water Lost}]{
		\includegraphics[width=3.25in]{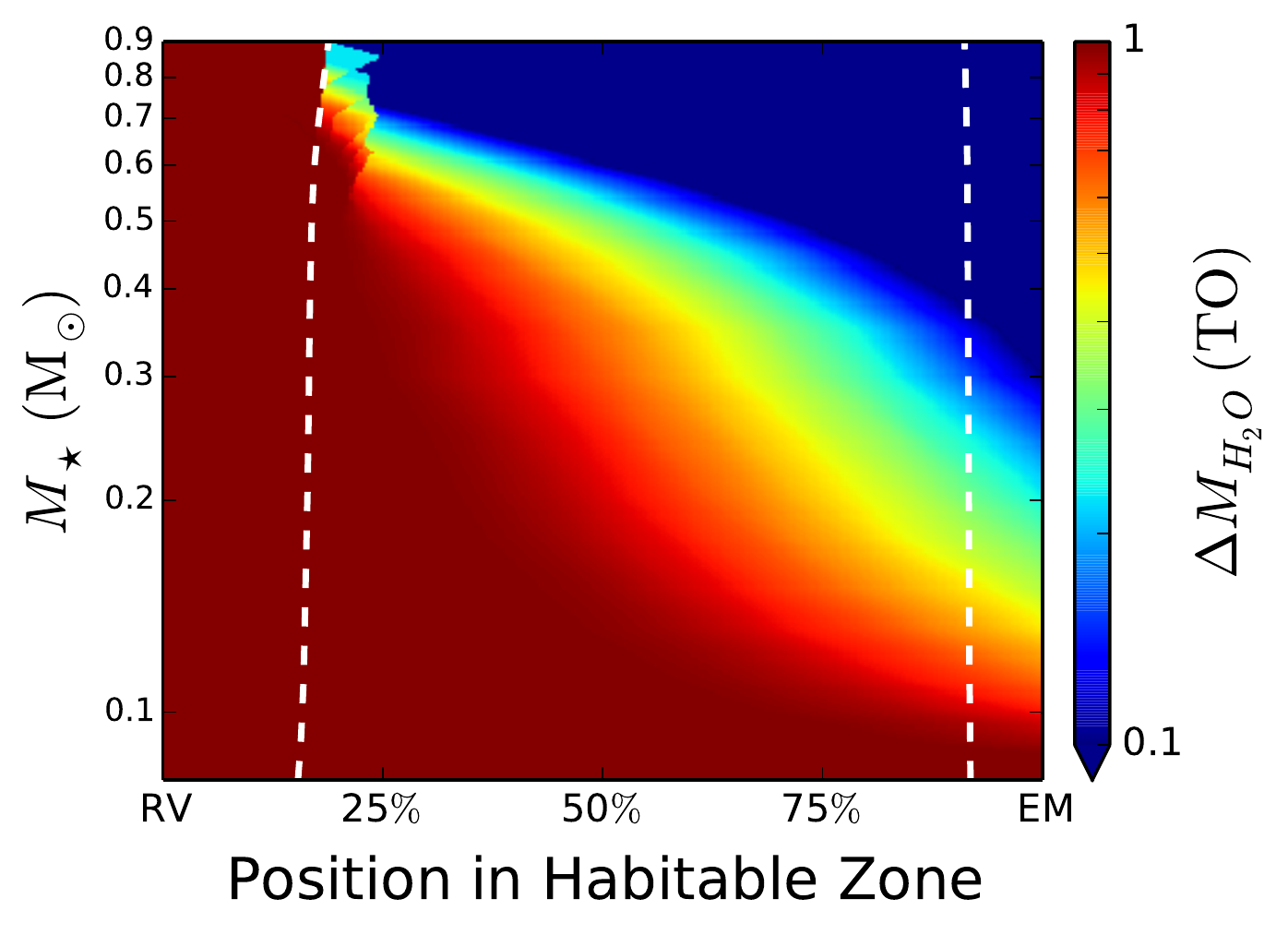}
	}
	\subfigure[\textbf{Oxygen in Atmosphere}]{
		\includegraphics[width=3.25in]{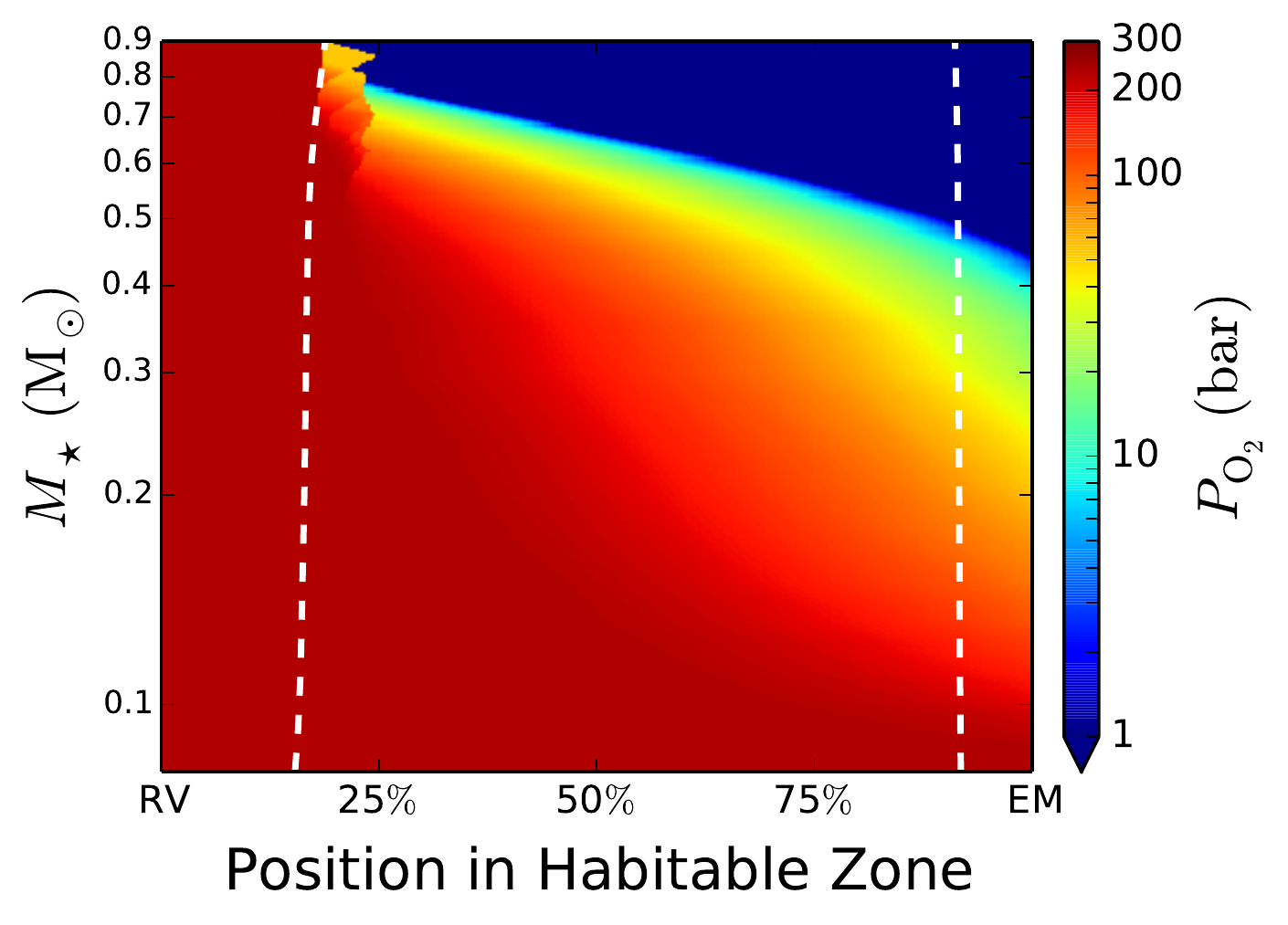}
	}
	\caption{Similar to Figure~\ref{fig:elim1TO} (1 TO), but assuming that the escape of hydrogen is diffusion-limited. This corresponds to planets with slow/ineffective oxygen sinks that retain all of the photolytically-produced O$_2$ in their atmospheres. While water loss amounts are generally lower than in the energy-limited case, planets throughout a large fraction of the HZ of M dwarfs are still desiccated. Moreover, the amount of oxygen that builds up is substantially greater than in Figure~\ref{fig:elim1TO}, since the oxygen cannot escape if the loss of hydrogen is diffusion-limited. Thus, planets that lose 1 TO of water build up $\frac{16}{18}\times 270 = 240$ bars of O$_2$ in their atmospheres.\vspace*{0.1in}
	\label{fig:dlim1TO}}
\end{figure*}

\begin{figure*}[t]
	\centering 
	\textbf{\large \sc Diffusion-Limited Escape: 1 $\mathrm{M}_\oplus$, 10 TO}\par\medskip 
	\subfigure[\textbf{Water Lost}]{
		\includegraphics[width=3.25in]{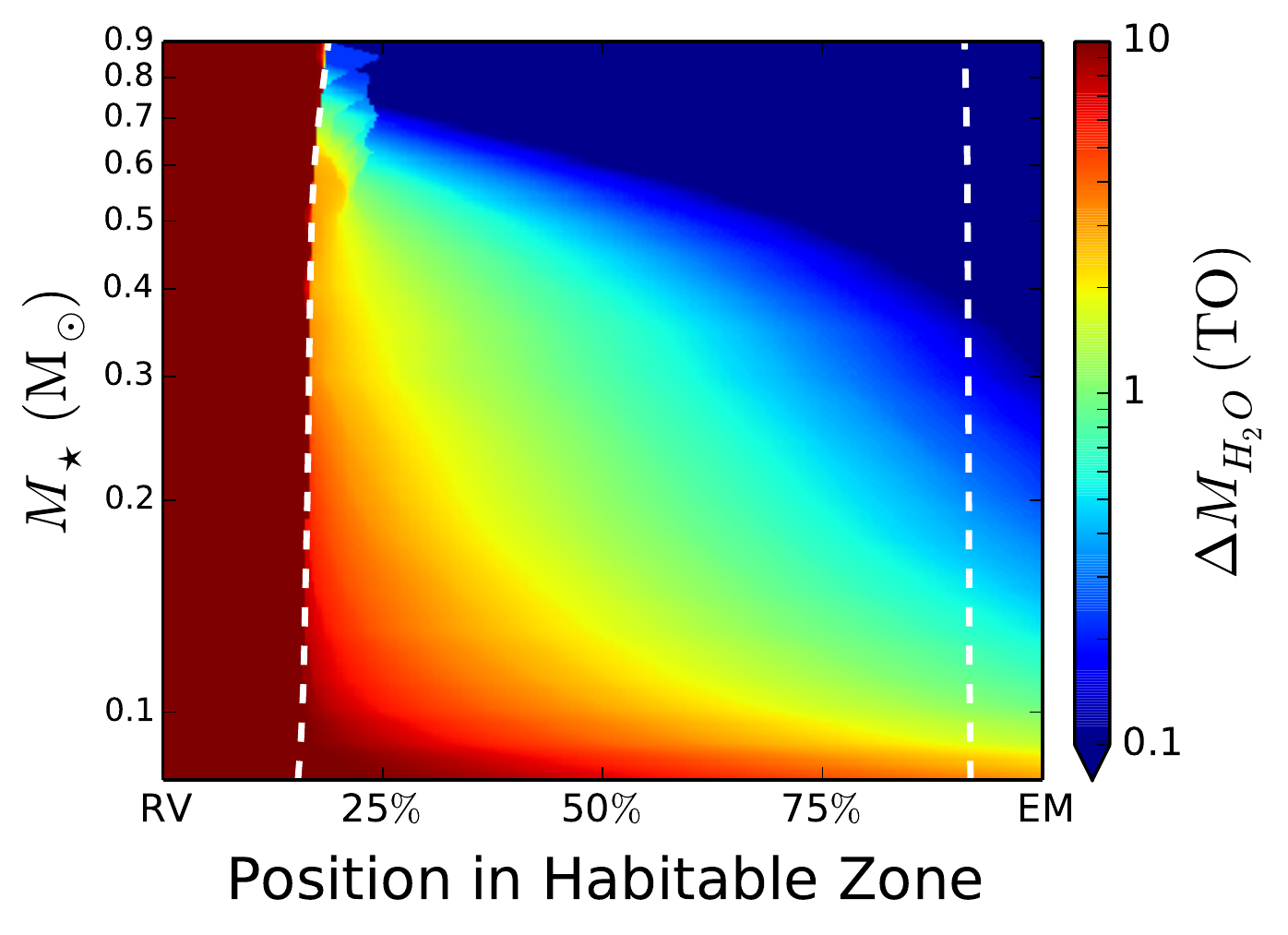}
	}
	\subfigure[\textbf{Oxygen in Atmosphere}]{
		\includegraphics[width=3.25in]{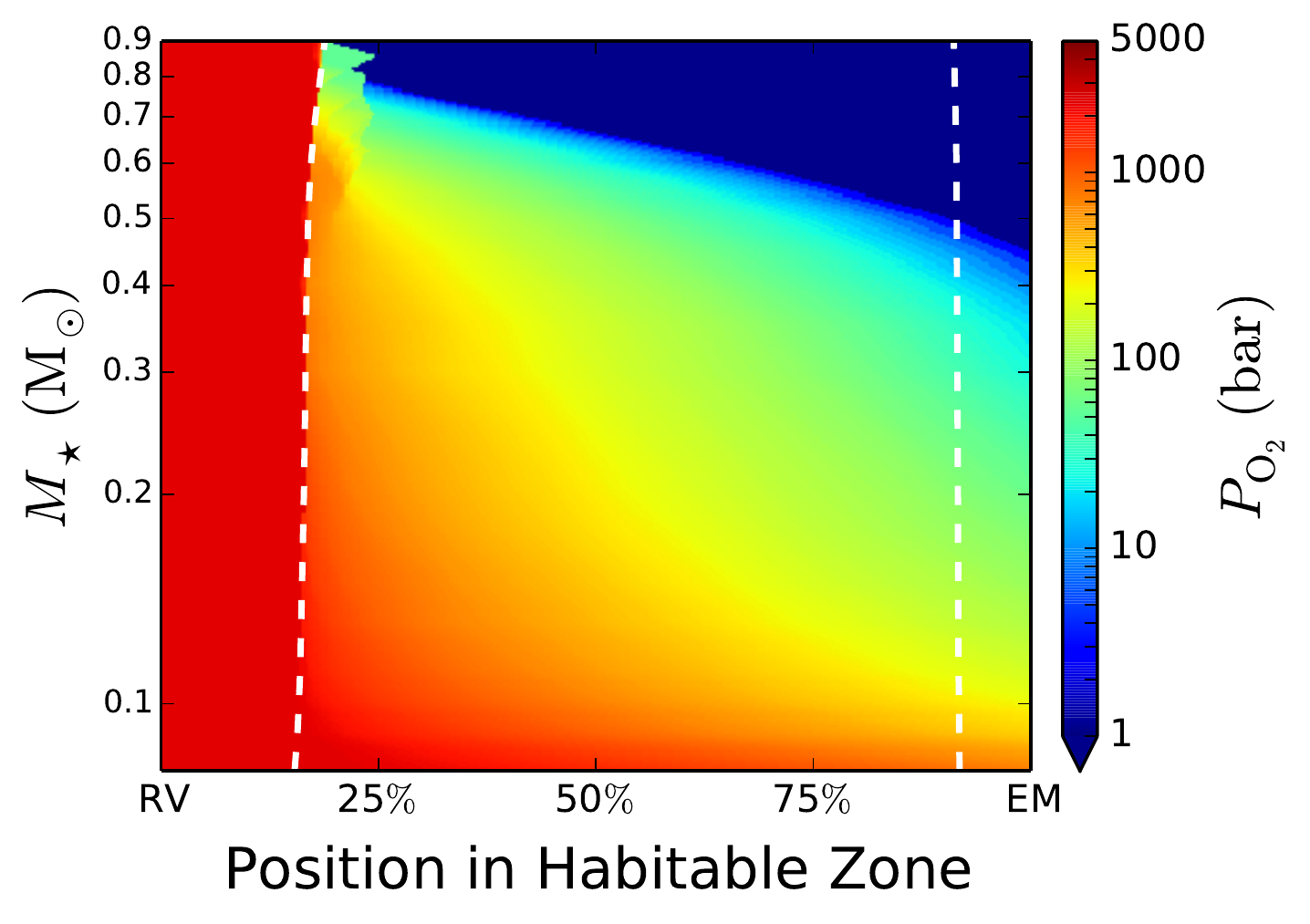}
	}
	\caption{Same as the previous figure (diffusion-limited escape), but for an initial water content of 10 TO; compare to Figure~\ref{fig:elim10TO}, the corresponding energy-limited case. Once again, water loss amounts are smaller, but oxygen amounts greater than 2000 bars are now possible around the lowest mass M dwarfs.\vspace*{0.1in}
	\label{fig:dlim10TO}}
\end{figure*}

In the previous section we assumed that the rate of oxygen removal at the surface was much higher than the rate at which oxygen was produced. We now repeat all calculations in the opposite limit, assuming the production rate is much higher than the absorption rate. As discussed in \S\ref{sec:o2atmospheres}, this could be the case for water worlds or for planets with a pre-oxidized surface/interior, inefficient outgassing of reducing compounds, inefficient resurfacing processes, etc. In these runs, the oxygen remains in the atmosphere and hydrogen must diffuse through it in order to escape. We therefore calculate loss rates in the diffusion limit as described in \S\ref{sec:difflim}.

Our results are plotted in Figures~\ref{fig:dlim1TO} and \ref{fig:dlim10TO} for initial surface water contents of 1 and 10 TO, respectively. Water loss amounts are generally slightly lower (compare to Figures~\ref{fig:elim1TO} and \ref{fig:elim10TO}), since the diffusion limit is slower than the energy-limited escape rate at early times. Nevertheless, planets throughout a large portion of the HZ of M dwarfs are still desiccated in the 1 TO case. Planets with larger water inventories lose $> 1$ TO around low mass M dwarfs and close to the inner edge of high mass M dwarfs.

In the panels on the right, we see that oxygen buildup is larger than in the energy-limited case; recall that these panels now represent the equivalent pressure of oxygen \emph{in the atmosphere} at the end of the runaway phase. Despite the subdued water loss, the fact that no oxygen escapes leads to the buildup of $\sim 240$ bars of O$_2$ (the equivalent pressure of O$_2$ in 1 TO of water) throughout a large portion of the HZ in Figure~\ref{fig:dlim1TO}. In Figure~\ref{fig:dlim10TO}, planets build up between $\sim 100$ and $\sim 1000$ bars of O$_2$ in their atmospheres throughout most of the HZ of M dwarfs. Note also that, unlike in the previous figures, the oxygen amount is a monotonic function of the position in the HZ; because no oxygen escapes, planets closer to the inner edge build up more O$_2$.
%
\begin{figure*}[t]
	\centering 
	\textbf{\large \sc Energy-Limited Escape: 5 $\mathrm{M}_\oplus$, 10 TO}\par\medskip
	\subfigure[\textbf{Water Lost}]{
		\includegraphics[width=3.25in]{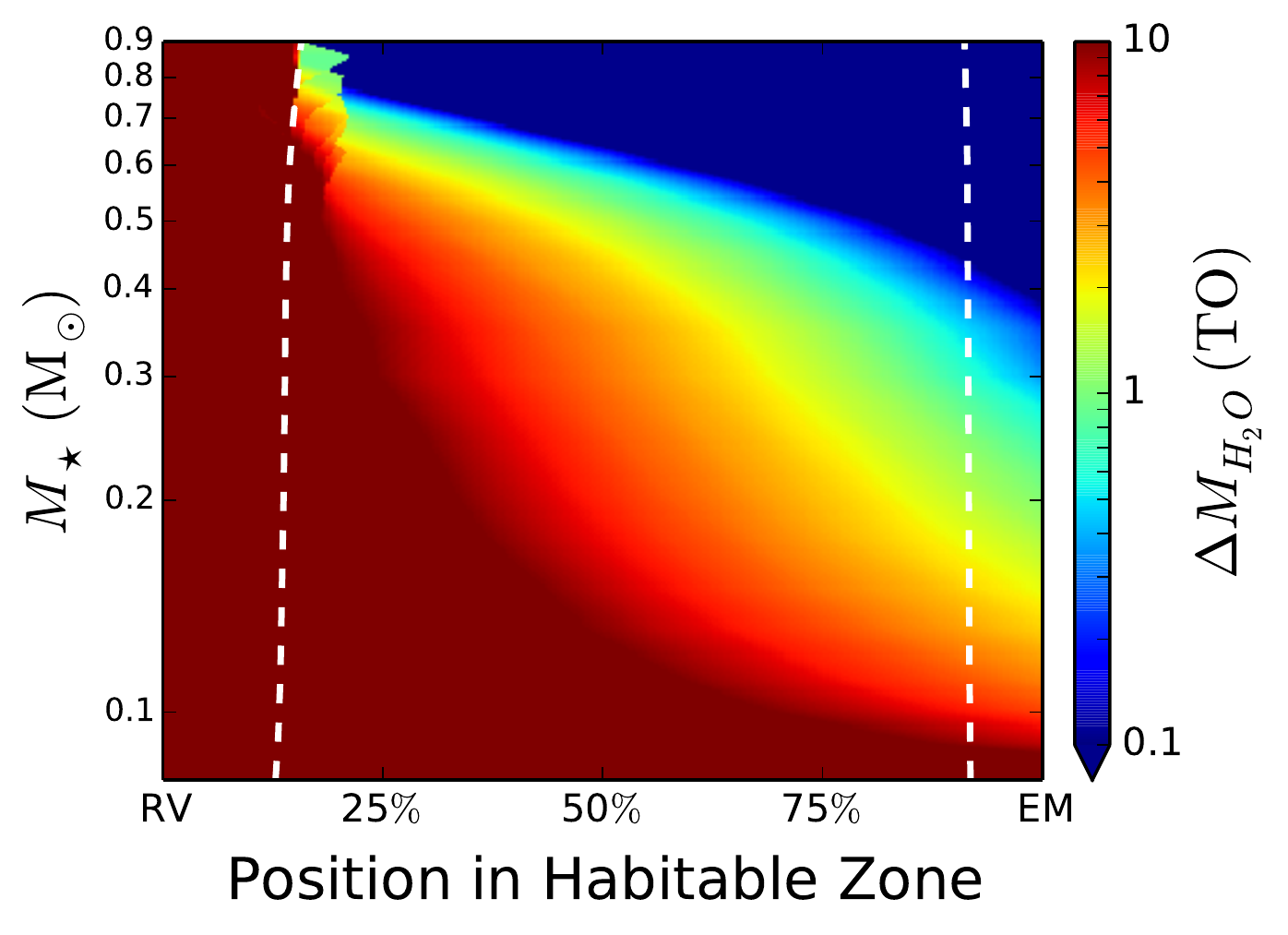}
	}
	\subfigure[\textbf{Oxygen Absorbed by Surface}]{
		\includegraphics[width=3.25in]{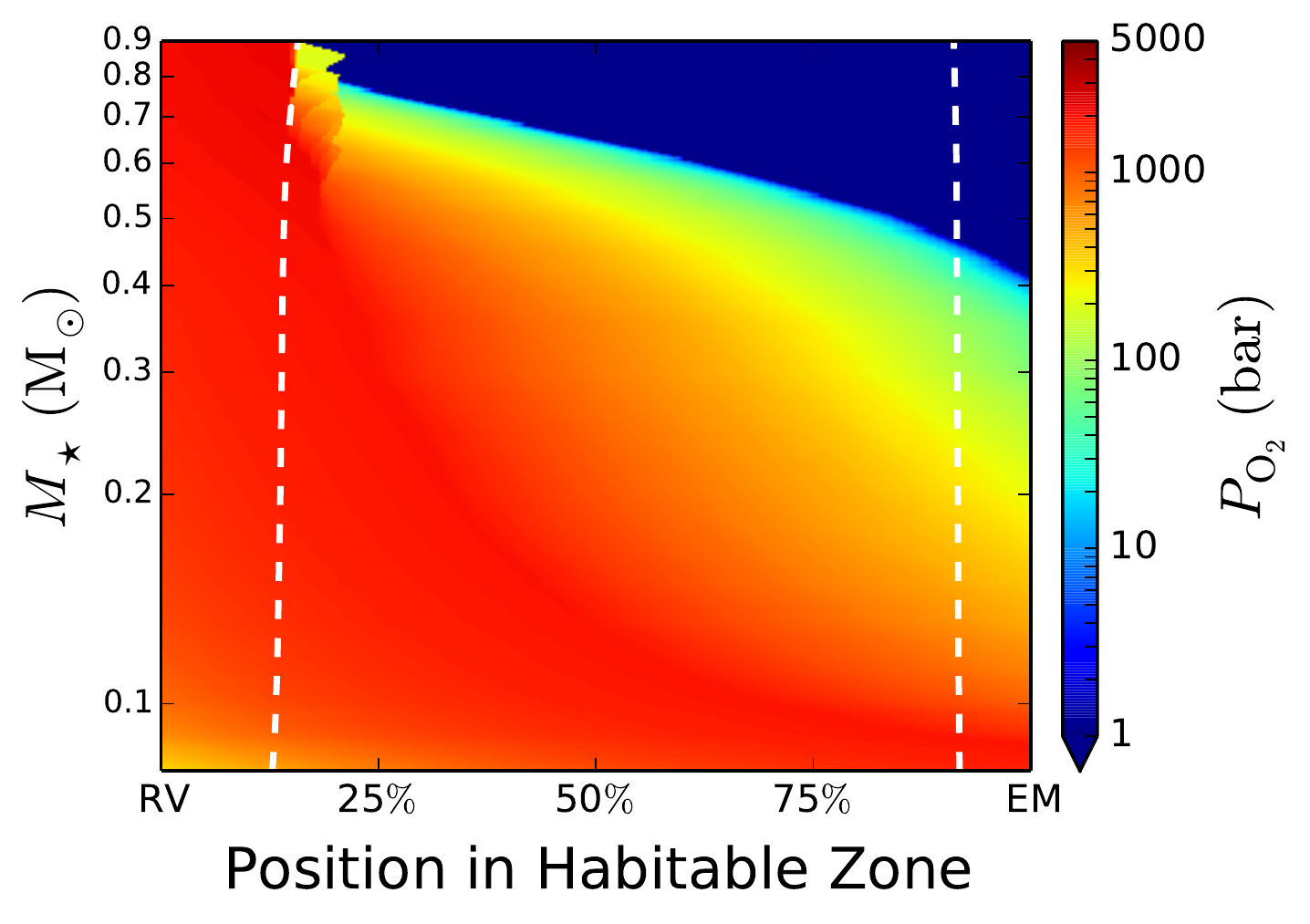}
	}
	\caption{Similar to the previous figures, but for a super-Earth of mass 5M$_\oplus$ with 10 TO, assuming energy-limited escape. Note the large fraction of the HZ in the lower left portion of the left panel where planets are completely desiccated. Elsewhere, super-Earths lose several TO of water. In the right panel, thousands of bars of O$_2$ are absorbed at the surface of planets throughout most of the HZ of M dwarfs.\vspace*{0.1in}
	\label{fig:elimSE}}
\end{figure*}
%
\begin{figure*}[t]
	\centering
	\textbf{\large \sc Diffusion-Limited Escape: 5 $\mathrm{M}_\oplus$, 10 TO}\par\medskip 
	\subfigure[\textbf{Water Lost}]{
		\includegraphics[width=3.25in]{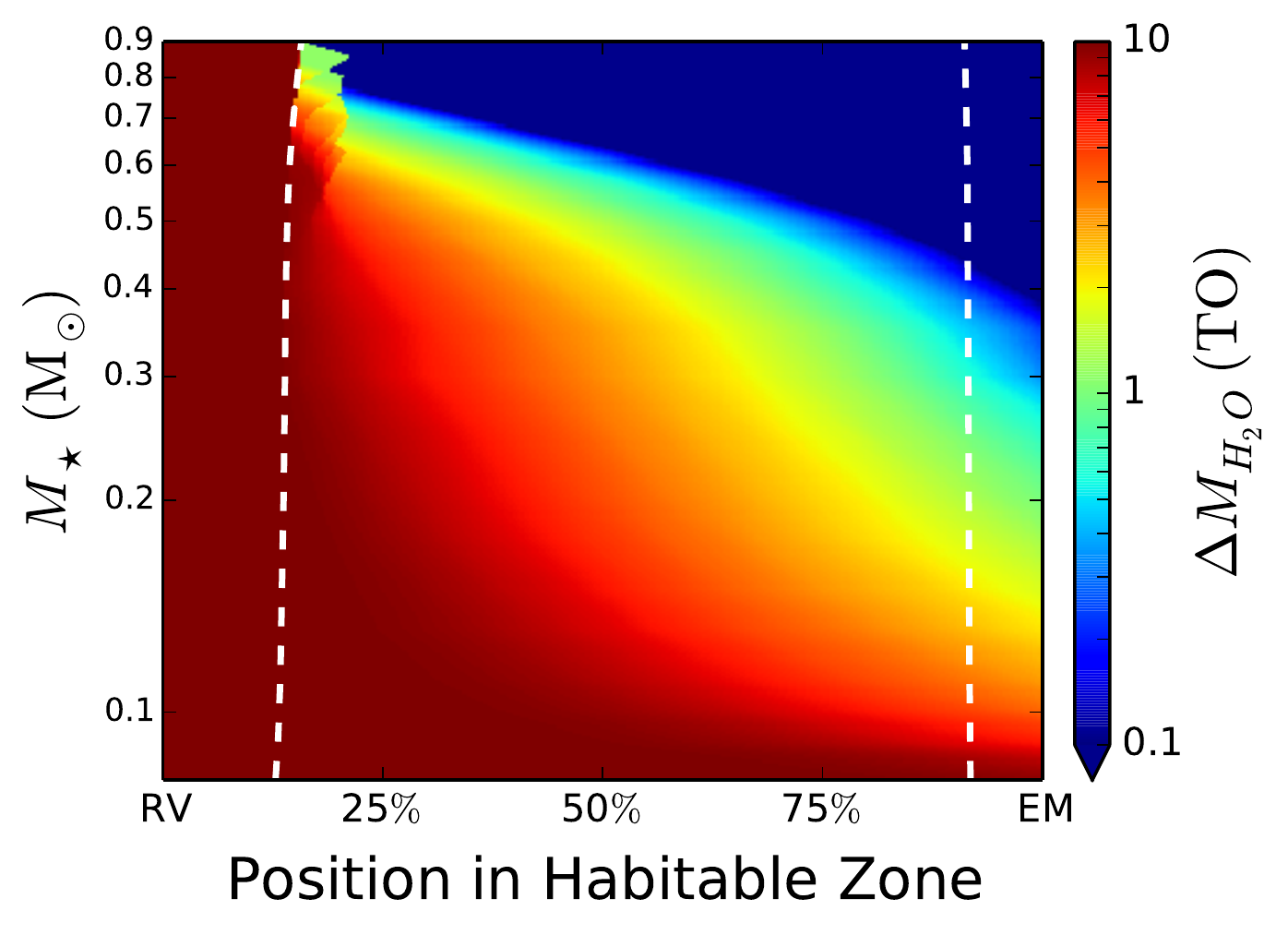}
	}
	\subfigure[\textbf{Oxygen in Atmosphere}]{
		\includegraphics[width=3.25in]{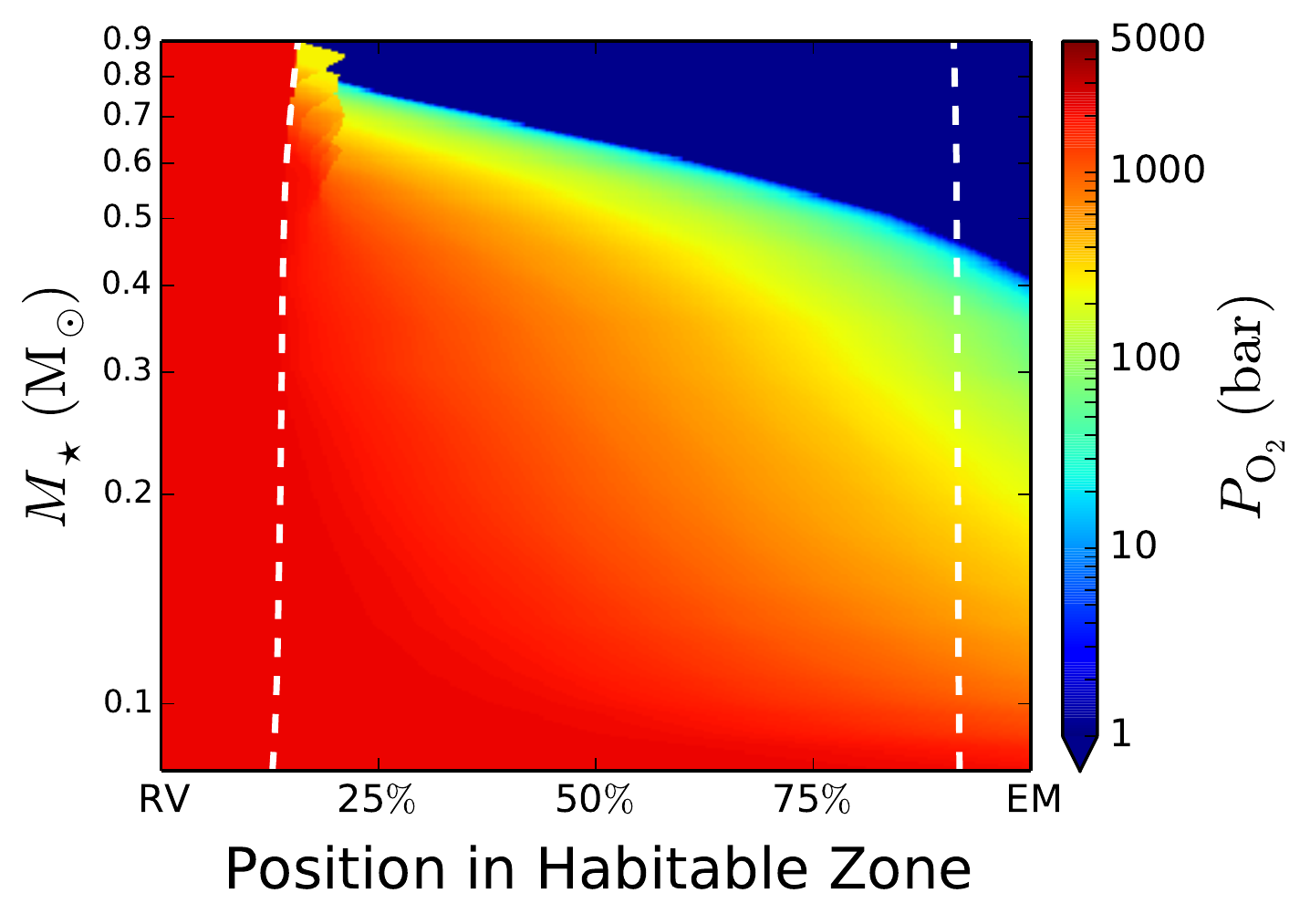}
	}
	\caption{Similar to Figure~\ref{fig:elimSE} (a 5 $\mathrm{M}_\oplus$ super-Earth with 10 TO of surface water), but assuming diffusion-limited escape. Despite a slight decrease in the total water loss amounts, the results are very similar to those in the energy-limited case. In general, these planets lose several TO of water and build up several hundred to a few thousand bars of O$_2$ in their atmospheres. \vspace*{0.1in}
	\label{fig:dlimSE}}
\end{figure*}
\subsection{Higher Planet Mass}
\label{sec:superearths}
In Figure~\ref{fig:elimSE} we show the results for a 5 M$_\oplus$ super-Earth with an initial water inventory of 10 TO and efficient oxygen sinks. Compared to a 1 M$_\oplus$ planet (Figure~\ref{fig:elim10TO}), both water loss and oxygen amounts are significantly \emph{higher}. Thousands of bars of O$_2$ are now retained throughout most of the HZ. Close to the inner edge of the lowest mass M dwarfs, planets with larger initial water inventories can lose several tens of TO.

This perhaps counter-intuitive result stems from the fact that the crossover mass (\ref{eq:mcfhref}) scales inversely with the surface gravity; on super-Earths, the escape of oxygen is greatly suppressed, leading to faster buildup at the surface. Moreover, when the oxygen does not escape, the loss of the ocean happens more quickly, since the escape of hydrogen is nine times more efficient at depleting the water content of the planet (at fixed $\mathcal{F}_\mathrm{XUV}$). Recall that in the limit $\eta \rightarrow 0$, (\ref{eq:mdotocean}) gives $\dot{m}_\mathrm{ocean} = 9\dot{M}_\mathrm{EL}$.

In Figure~\ref{fig:dlimSE}, we perform the same calculation, but for diffusion-limited escape, assuming the O$_2$ remains in the atmosphere. Once again, both the amount of water lost and the oxygen pressure are substantially higher than in the 1 M$_\oplus$ case. This is a straightforward consequence of (\ref{eq:difflim}); the diffusion limit flux scales inversely with the scale height of the background atmosphere, which is smaller on the super-Earth by a factor of $\sim$ 2.2 due to the higher surface gravity.

Perhaps even more interestingly, Figures~\ref{fig:elimSE} and ~\ref{fig:dlimSE} are very similar; the desiccation process of super-Earths is relatively insensitive to the escape regime and to whether or not the oxygen remains in the atmosphere or is absorbed by the surface. This similarity is due to the fact that the energy-limited escape rate and the diffusion-limited escape rate are comparable for the range of XUV fluxes received by these planets. This applies even at early times, when the XUV flux is very high; the escape of even a small amount of oxygen in the energy-limited regime tends to slow down the rate of ocean loss, given that a large fraction of the XUV energy goes into driving the escape of the heavier species.

It is important to note that since we have been plotting the O$_2$ equivalent \emph{pressure} rather than the actual amount, we must take care in comparing it between planets of different masses. Under the plane-parallel approximation, the atmospheric pressure scales as $M_\mathrm{p}/R_\mathrm{p}^4$. Given that a 5$\mathrm{M}_\oplus$ planet with an Earth-like composition has $R_\mathrm{p} = 1.52\mathrm{R}_\oplus$ \citep{FMB07}, the pressure exerted by a fixed amount of O$_2$ will actually be \emph{smaller} (by about 6\%) on the higher mass planet. Nevertheless, we see from these figures that the O$_2$ pressures are a factor of $2-3$ times \emph{larger} on a super-Earth than on an Earth-mass planet, implying substantially more O$_2$ buildup. 

\begin{figure}[h]
  \begin{center}
    \leavevmode
      \psfig{file=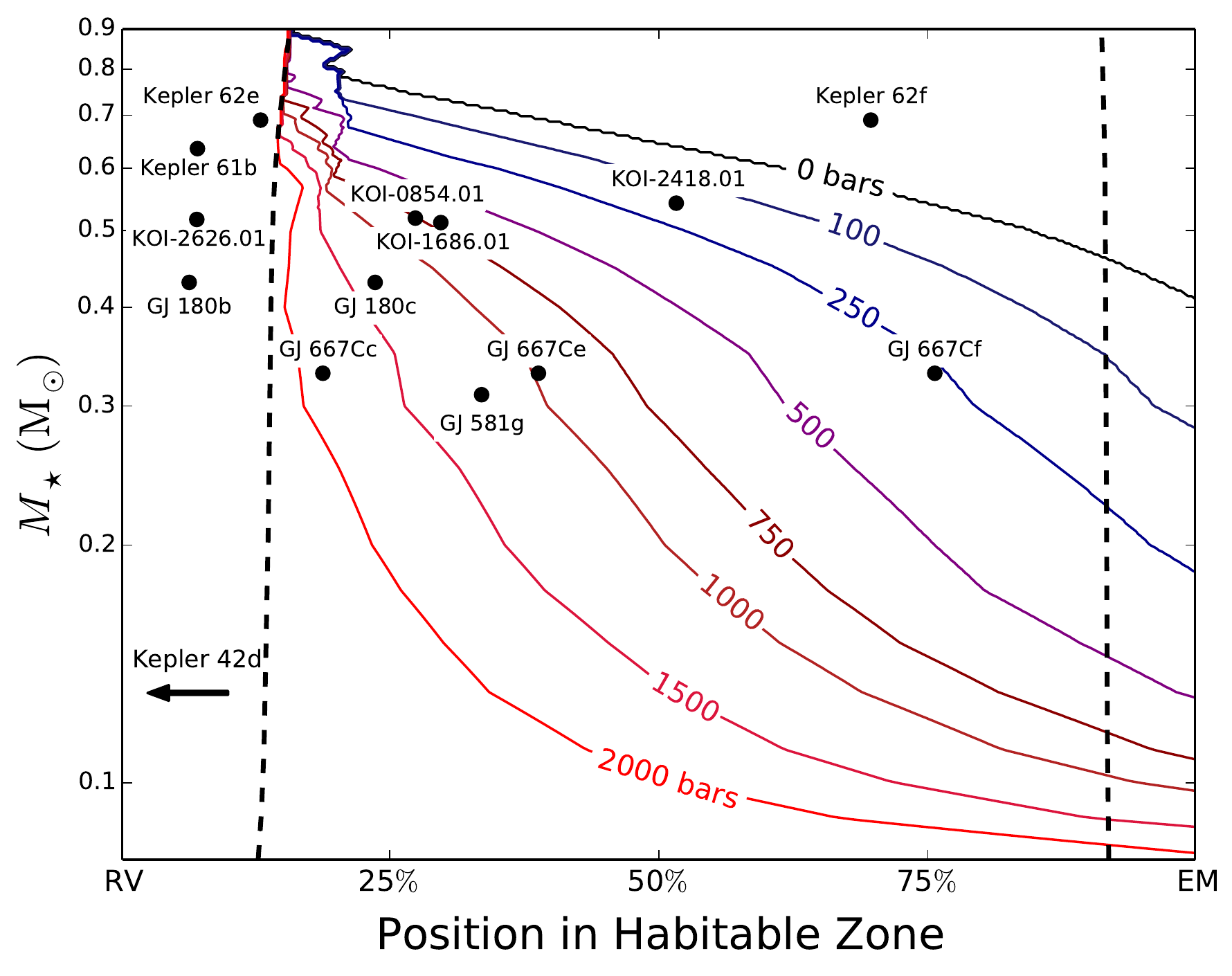,width=3.25in}
       \caption{A selection of M and K dwarf super-Earths that could have detectable O$_2$ atmospheres if they formed with abundant surface water; as before, the dashed lines represent the RG and MG limits. Calculations were performed assuming a mass of 5 M$_\oplus$, an initial water content of 10 TO, and diffusion-limited escape. Contours correspond to the equivalent O$_2$ atmospheric pressure in bars at the end of the runaway phase, assuming all of the O$_2$ remains in the atmosphere. Of the planets shown here, only Kepler 62f does not build up any oxygen.\vspace*{0.1in}}
     \label{fig:exoplanets}
  \end{center}
\end{figure}

For reference, in Figure~\ref{fig:exoplanets} we overplot a handful of known planets on the oxygen pressure contours of Figure~\ref{fig:dlimSE} (b): GJ 581g\footnote{The existence of GJ 581g has recently been contested; see \cite{ROB14}.} \citep{VOG10}, Kepler 61b \citep{BAL13}, Kepler 62e \& f \citep{BOR13}, GJ 180c \citep{TUO14}, the GJ 667 system \citep{ANG13}, and four Kepler candidate planets \citep{DC13}; Kepler 42d \citep{MUI12}, which is interior to the HZ, is indicated by an arrow. As in Figure~\ref{fig:dlimSE}, we assume that the planet mass is 5 M$_\oplus$, the oxygen remains in the atmosphere, and the escape is diffusion-limited. Provided they formed with abundant water, many of the currently known super-Earths could have built up hundreds to thousands of bars of O$_2$. In particular, this could be the case for GJ 667Cc, which could have lost as many as 10 TO early on, accumulating close to 2000 bars of O$_2$; as a result, it may not be habitable today.

Whether this oxygen remains in these planets' atmospheres past the early runaway phase depends on the efficiency of their surface sinks. Given poor constraints on exoplanet tectonics, the physics of oxygen absorption by a magma ocean, and other aspects of exoplanet atmospheres, it is reasonable to expect that some super-Earths may be unable to remove all the photolytically-produced oxygen within the ages of their systems.

\subsection{The Rate of Oxygen Buildup}
\label{sec:constantpodot}
While the final O$_2$ equivalent pressure is a complex function of the stellar/planetary mass and the semi-major axis, our results in the energy-limited regime can be understood in fairly simple terms by considering the mass loss rates (\ref{eq:mdothup})-(\ref{eq:mdotodown}) derived in Appendix A. In particular, the rate of oxygen accumulation (\ref{eq:mdotodown}) is a function of both $\eta$ and $\dot{M}_\mathrm{EL}$, which are themselves functions of the XUV flux. By expressing $\dot{M}_\mathrm{EL}$ as a function of $\eta$, we show in Appendix B that the rate at which oxygen accumulates in the atmosphere/at the surface is completely independent of the XUV flux above the critical value given in (\ref{eq:fxuvmin}). This rate is
\begin{align}
\label{eq:constantpodot}
\dot{P}_\mathrm{O_2} = 5.35 \left( \frac{M}{\mathrm{M_\oplus}} \right)^2	\left( \frac{R}{\mathrm{R_\oplus}} \right)^{-4}\ \mathrm{bars\ Myr^{-1}}.
\end{align}
When $X_\mathrm{O}/X_\mathrm{H} = 1/2$, this is also the rate at which oxygen builds up in the atmosphere in the diffusion-limited regime. In the diffusion limit, this expression holds at early times, when the atmosphere is still predominantly H$_2$O; towards the end of the escape regime, the rate at which oxygen builds up tapers off due to the decreased H escape flux.

This result may seem very counter-intuitive, since it implies that the rate of oxygen buildup is independent of the XUV flux and relatively independent of the escape regime. In particular, one might expect that in the energy-limited regime, an increase in the XUV flux would lead to more oxygen escape and thus a slower rate of buildup. However, increasing $\mathcal{F}_\mathrm{XUV}$ also leads to a higher \emph{hydrogen} escape rate and a faster net production of O atoms. While the ratio of the oxygen to hydrogen mass escape rates in (\ref{eq:moupmhup}) approaches a maximum value of 8 for $\eta \rightarrow 1$, the \emph{difference} between the oxygen production and escape rates remains constant. 

The constant rate of O$_2$ buildup is a straightforward consequence of mass fractionation during hydrodynamic escape. In Appendix B we show that (\ref{eq:constantpodot}) corresponds to a flux of oxygen atoms into the atmosphere equal to 
\begin{align}
F_\mathrm{O}^\mathrm{atm} = 5bgm_H/kT,
\end{align}
which is precisely the diffusion limit flux for atomic oxygen through a background of atomic hydrogen. Put another way, in order for oxygen to be retained after photolysis, it must diffuse out of the hydrodynamic flow that is dragging it away, and the rate at which it can do so is equal to the diffusion limit. Therefore, regardless of whether the escape of \emph{hydrogen} is energy-limited or diffusion-limited, oxygen buildup will occur at its diffusion limit.

Since this rate is constant in the energy-limited regime and declines slowly in the diffusion-limited regime, the primary factor controlling the final oxygen amount retained by the planet is the duration of the hydrodynamic escape phase. In general, an Earth-mass planet will build up oxygen at a constant rate of about 5 bars/Myr until (a) the planet leaves the runaway regime (enters the HZ); (b) enough O$_2$ builds up in the atmosphere to slow the H escape rate; (c) the planet loses all surface water; or (d) the XUV flux drops below $\mathcal{F}_\mathrm{crit}$.

\begin{figure}[h]
	\centering   
	\includegraphics[width=3.25in]{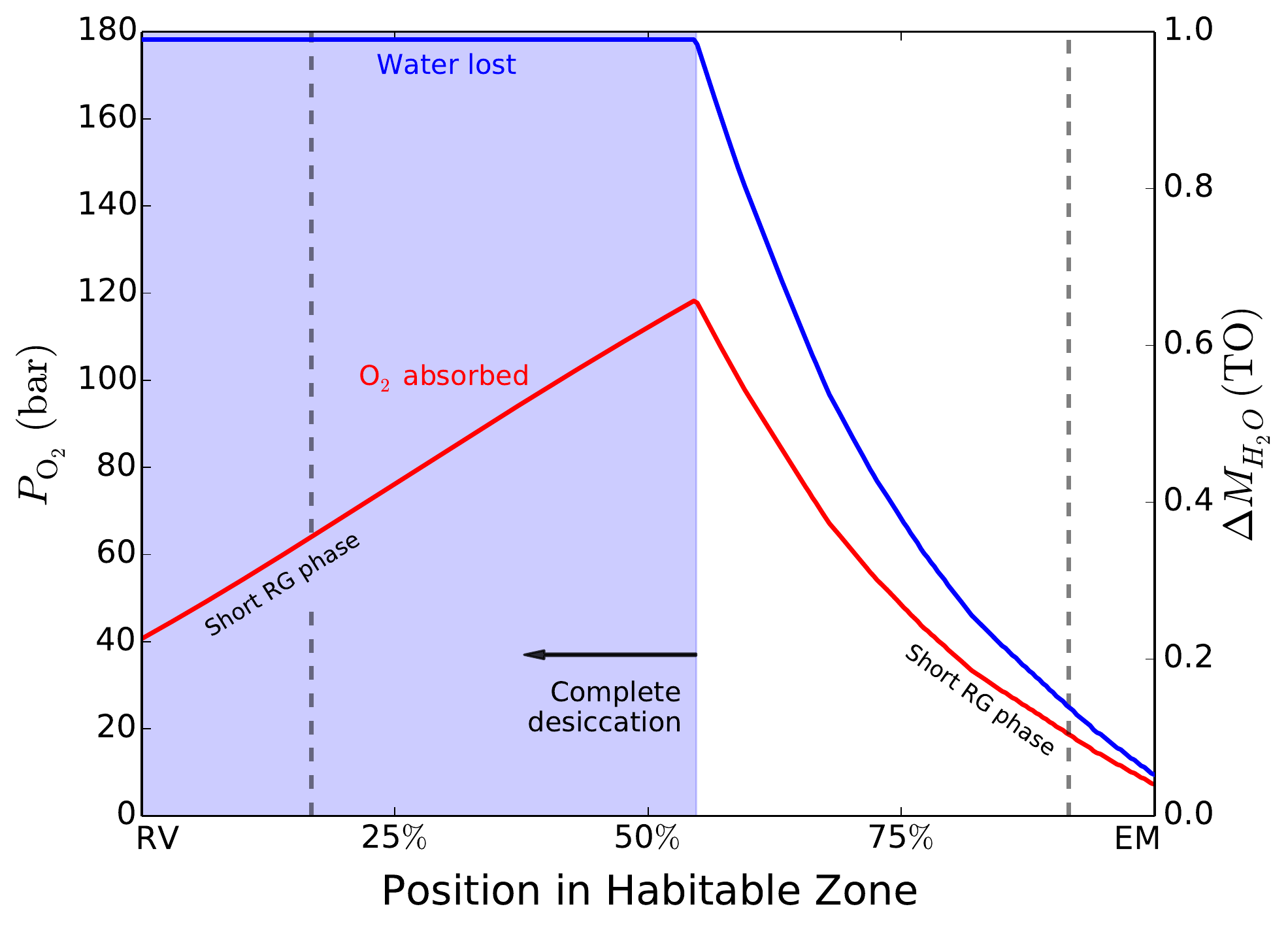}
	\caption{Cross-section along $M_\star = 0.4 \mathrm{M_\odot}$ in Figures~\ref{fig:elim1TO} (a) and (b), showing the oxygen absorbed by the surface (red) and total amount of water lost (blue) as a function of the position in the HZ. Water loss scales with the time spent in the runaway phase; interior to a critical distance, complete desiccation occurs (shaded region). From right to left, the O$_2$ amount initially increases due to the increase in the amount of water lost. However, interior to the critical distance, complete desiccation occurs at progressively earlier times. The higher XUV flux early on results in more oxygen escape and less buildup. See text for a discussion.\vspace*{0.1in}
	\label{fig:xsect}}
\end{figure}

This helps shed light on the behavior seen in several of the figures in the previous sections. Consider, for instance, Figure~\ref{fig:elim1TO}, where the final oxygen pressure is not a monotonic function of either $M_\star$ or the position in the HZ, peaking close to the outer edge of the HZ in some cases. This trend is shown more clearly in Figure~\ref{fig:xsect}, where we plot a cross-section along $M_\star = 0.4 \mathrm{M_\odot}$ in Figures~\ref{fig:elim1TO}~(a) and (b). The blue curve corresponds to the amount of water lost and the red curve corresponds to the final oxygen pressure. 

Since $M_\mathrm{p} = 1 \mathrm{M_\oplus}$ and the escape regime is energy-limited, all planets build up O$_2$ at the same rate of 5.35 bars/Myr. The difference in the final value of $P_\mathrm{O_2}$ is solely due to the duration of the escape. Towards the outer edge of the HZ, planets experience short-lived runaway greenhouse phases; see Figure~\ref{fig:gtime} (b). As one moves from the outer HZ to the inner HZ, the increasing duration of the RG phase leads to higher O$_2$ pressures.

However, at a certain critical position in the HZ ($\sim$ 55\% in Figure~\ref{fig:xsect}), the final O$_2$ pressure begins to decrease. This is because planets in the shaded region are completely desiccated, which terminates the escape phase and sets an upper limit to the O$_2$ pressure. Interior to this distance, faster water loss (blue line) results in an earlier desiccation time, leading to lower O$_2$ amounts.

We note, finally, that the rate of O$_2$ buildup is constant only when the XUV flux exceeds $\mathcal{F}_\mathrm{crit}$. Below this flux, the escape of hydrogen is energy-limited, the oxygen escape rate is zero and its rate of buildup is linearly proportional to the H escape flux. However, nearly all Earth-mass planets in our runs enter the HZ before the XUV flux drops below the critical value. Compare Figures~\ref{fig:xuvevol} and \ref{fig:bolfluxevol}, noting that planets around M dwarfs remain in a runaway for $0.1 - 1$ Gyr, during which time $\mathcal{F}_\mathrm{XUV} \gtrsim 300\mathrm{\mathcal{F}_\oplus}$ for a saturation time of 1 Gyr. For a shorter saturation time (typical of K and G dwarfs) and/or a higher planet mass (for which the critical flux is higher), planets may leave the oxygen escape regime prior to the end of the runaway greenhouse.

\section{Discussion}
\label{sec:discussion}
\subsection{Prolonged Runaway Greenhouse}
\label{sec:prolonged}
We showed in the first part of this paper that the pre-main sequence evolution of M dwarfs causes their habitable zones to migrate inwards by nearly an order of magnitude in semi-major axis during the first few hundred Myr. This means that many of the planets currently in the HZs of M dwarfs were not in the HZs when they formed. The fact that these planets spend several tens to several hundreds of Myr in insolation-induced runaway greenhouses establishes initial conditions in stark contrast to those on Earth, the only known habitable planet to date. This could significantly compromise their habitability.

Because of the extended runaway greenhouse, the atmospheric and thermal evolution of many M dwarf planets could follow very different paths than Earth's. Chemical equilibrium between the atmosphere and a magma ocean during a runaway could lead to the buildup of hundreds of bars of CO$_2$ early on \citep{ELK08}. High surface temperatures could also inhibit plate tectonics by promoting rapid lithospheric healing and grain growth, increasing the viscosity and erasing weak zones where plate subduction can occur \citep[e.g.,][]{DRI13}. This may prevent the onset of a carbonate-silicate cycle on these planets, making them unable to remove atmospheric CO$_2$ and maintaining permanently high surface temperatures.

Currently, the closest analog to the planets we consider here is Venus, which as a result of a prolonged runaway greenhouse phase underwent complete desiccation, irreversible crustal oxidation, termination of plate tectonics, and the buildup of a dense CO$_2$ atmosphere which maintains the surface temperature over 700 K. Whether this is the fate of all planets that undergo prolonged runaway greenhouses is unclear. Earth itself is thought to have been in an impact-induced runaway greenhouse for a few Myr \citep{ZAH88,HAM13}, yet it is habitable today. Detailed models coupling the atmosphere to the interior evolution of planets around M dwarfs should be performed in the future in order to address the long-term consequences of the early RG phase.

\subsection{The Fate of the Oxygen}
\label{sec:oxygenfate}
In the second part of the results, we showed that the prolonged runaway greenhouse can lead to the loss of several TO of water and the buildup of hundreds to thousands of bars of O$_2$ on M dwarf planets that end up in the HZ. We considered two limiting cases regarding the escape: energy-limited and diffusion-limited. We argued that the former applies to planets with efficient surface sinks for the O$_2$, as the sinks prevent oxygen from becoming a major atmospheric gas and thus bottlenecking the escape of hydrogen via diffusion. The latter applies to the opposite case, where the O$_2$ builds up quickly and caps the hydrogen escape rate at the diffusion limit.

In reality, a planet with an Earth-like redox state is likely to start out in the energy-limited regime and transition to the diffusion-limited regime as its surface sinks get overwhelmed; our results for water loss and O$_2$ buildup should therefore bracket these processes on many M dwarf planets. In fact, any Earth-mass planet that removes oxygen at a rate slower than $\sim 5$ bars/Myr ($\sim 25$ bars/Myr for a super-Earth) will build up O$_2$ in its atmosphere during the runaway phase and likely transition to diffusion-limited escape. Moreover, \emph{any} major atmospheric gas will tend to cap the escape at the diffusion limit. If species such as CO$_2$ or N$_2$ are abundant relative to H$_2$O during the runaway, these would also enforce diffusion-limited escape.

Nevertheless, we have shown that even at the diffusion limit, extreme water loss and atmospheric O$_2$ buildup may occur on these planets. Given the ability of surface processes to remove tens to hundreds of bars of O$_2$ per Gyr (\S\ref{sec:o2atmospheres}), we might expect that some planets that build up oxygen atmospheres in excess of $\sim$ 100 bar could retain detectable amounts of O$_2$ today. Such elevated quantities of O$_2$ are possible throughout the HZ of all M dwarfs, except near the outer edge of those more massive than about 0.5 M$_\odot$, where planets are in runaway greenhouses for only a few Myr. 

It is important to note that, because of the potential volatile deficiency of planets that form \emph{in situ} in the HZs of M dwarfs \citep{RAY07,LIS07}, many of the planets in the HZ with abundant surface water early on could be different from Earth in composition, given that they may have formed beyond the snow line and migrated in. In particular, they could have large inventories of highly reducing compounds, since hydrogen-rich species such as methane and ammonia condense at large distances from the host star and are thus easily accreted during planet formation. Not only might these planets have highly reducing surfaces, but outgassing of reduced compounds could also lead to the quick removal of atmospheric O$_2$. This is especially the case for planets with nonzero eccentricity around low mass M dwarfs, as strong tidal forces in the HZ can drive vigorous volcanism \citep{JBG08,BAR13}. On the other hand, whether these planets develop a magma ocean capable of absorbing most of the O$_2$ is unclear, as they may not have rocky surfaces, but instead a thick layer of water and ice extending down to great depths.
 
Another issue is the possibility that planets forming beyond the snow line could develop substantial hydrogen/helium envelopes before they migrate into the HZ. While the hydrogen may act as a direct sink for the oxygen (via the formation of H$_2$O), the envelope could also inhibit the escape of water in the first place. A dense enough envelope could effectively shield the surface and prevent the dissociation of H$_2$O, provided the water settles below the base of the hydrodynamic wind. However, planetary cores that accrete dense H/He envelopes may be unable to completely shed them, in which case the planet will never be terrestrial (and probably not habitable). In order for the envelope to prevent water loss and O$_2$ buildup and yet not preclude the planet's eventual habitability, it must bear the brunt of the XUV flux early on and dissipate before the flux drops too quickly and atmospheric escape becomes negligible. We explore this possibility in detail in \cite{LUG14}, noting that such evaporated cores could potentially be habitable.

As for planets that happen to form \emph{in situ} with abundant water, one of the major factors controlling whether or not a long-term O$_2$ atmosphere will develop could be the interaction with a potential magma ocean (\S\ref{sec:magmaocean}). \cite{GIL09} and \cite{HAM13} argue that this process could be responsible for the lack of oxygen in the Venusian atmosphere today. It is entirely plausible that a deep magma ocean could remove most of the atmospheric O$_2$ on M dwarf planets, provided solidification of the mantle occurs only at the end of the runaway. If, on the other hand, solidification occurs after $\sim$ 10 Myr as it likely did on Venus \citep{LEB13}, planets that spend longer than this amount of time in a runaway may accumulate a large fraction of the O$_2$ in their atmospheres. Future work will investigate the mantle cooling process during runaway greenhouses on M dwarf planets.

\subsection{Implications for Habitability}
\label{sec:habitability}
We have shown that during the early high luminosity phase of M dwarfs, terrestrial planets can lose several Earth oceans of water. Over 10 oceans may be lost close to the inner edge of the HZ, particularly for super-Earths, whose hydrogen diffusion limit is higher, and for planets with efficient O$_2$ absorption processes, as these could prevent atmospheric O$_2$ from building up, resulting in an escape rate that approaches the energy-limited rate. Given the fundamental importance of water to life as we know it, complete planetary desiccation could severely hamper the ability of life to originate and evolve on planets around M dwarfs. Moreover, water is an essential ingredient for both plate tectonics \citep{MAC98,MS98} and the carbonate-silicate cycle, which together regulate CO$_2$ in the Earth's atmosphere. Without a mechanism to remove atmospheric CO$_2$, desiccated planets may build up dense CO$_2$ atmospheres and maintain high surface temperatures even after the end of the greenhouse phase, much like Venus. In this case, even a late delivery of water by comets or asteroids may be unable to restore the planet's habitability. 

We have also shown that planets that lose significant amounts of water retain tens, hundreds, or even thousands of bars of photolytically-produced O$_2$, which in certain cases could remain in the atmosphere. It is widely accepted that prebiotic chemistry happened in a reducing environment \citep{OPA24,HAL29}; moreover, life on Earth evolved in the absence of atmospheric oxygen for at least 1 Gyr \citep[e.g.,][]{SCH07,ANB14}. Early organisms relied on the free energy available in redox reactions involving a variety of hydrogen compounds; on an O$_2$-rich planet, organisms would have to compete with the oxygen for this free energy. Abiogenesis in the presence of massive amounts of atmospheric oxygen could therefore be difficult, though these issues should be investigated further.

\subsection{Other Remarks}
\label{sec:otherremarks}
We assumed that terrestrial planets form with abundant surface water. Planets that form dry and receive water at a late stage (such as by cometary impacts) are naturally more robust against water escape and oxygen buildup, since the runaway greenhouse will not last as long. The inner edge of the HZ for dry planets may also be significantly closer to the star \citep{ABE11}. Moreover, we only considered planets that migrate via disk interactions. Planet-planet scattering processes, for instance, need not occur early; if a terrestrial planet scatters into the HZ after the star settles onto the MS, it will be safe from the early runaway period.

However, we only considered water loss and O$_2$ buildup during the runaway greenhouse state. Water loss may occur during a moist greenhouse state \citep{KOP13} at the end of the star's contraction phase. Moreover, diffusion-limited escape of hydrogen can occur even on a planet that is not in a runaway state. \cite{WP14} show that an N$_2$-poor Earth can lose up to 0.3 TO and produce 66 bars of O$_2$ due to diffusion-limited escape. A planet whose crust and mantle have been highly oxidized during the runaway regime may be unable to remove this oxygen, leading to an O$_2$-rich atmosphere even if all the oxygen produced during the runaway was absorbed into the magma.

We neglected the possibility of cold-trapping of the water at the end of the runaway phase. On Earth, water vapor is strongly inhibited from reaching the stratosphere by condensation in the troposphere. However, due to the high surface temperature on a runaway planet, the thermal structure follows a dry adiabat throughout most of the troposphere \citep[e.g.,][]{KAS88}, along which the vapor pressure is lower than the saturation vapor pressure and thus water cannot condense. Nevertheless, as a planet gets desiccated and H$_2$O becomes progressively less abundant relative to O$_2$, a cold trap could eventually be established. In principle, this could prevent planets from becoming completely desiccated, allowing a small fraction of the initial water content to remain after the end of the runaway.

We also ignored atmospheric loss due to flares and stellar winds, which is likely significant around M dwarfs \citep{SCA07,KIS13} and can lead to the loss of substantially more water than we calculate here. Flares could also remove some of the oxygen, lowering the amount that builds up in the atmosphere or at the surface, particularly around the lowest mass M dwarfs. On the other hand, stars less massive than about 0.1M$_\odot$ may be in a supersaturation regime early on, saturating at XUV fluxes one or even two orders of magnitude below those of higher mass M dwarfs \citep{COO14}. This could reduce water loss rates from planets around these lowest mass stars.

We further assume blow-off conditions for oxygen are not met. \cite{TIA09} showed that super-Earths with CO$_2$ atmospheres receiving up to 1000 F$_\oplus$ are stable to carbon escape; oxygen escape may be similarly inhibited. However, as \cite{LAM11a} point out, this may not be the case on Earth-mass planets that lack IR-cooling species. On such planets, significantly less oxygen could build up in the atmosphere.

We only presented figures where the runaway greenhouse occurs interior to the RG limit. We also considered the case of planets that are in a runaway only while they are interior to the RV limit (the inner edge of the empirical HZ). We find that total water loss amounts and O$_2$ pressures are similar, but all contours shift to the left (following the shift in the HZ boundary). This is due to the fact that these planets remain in a runaway for less time.

All plots shown in the paper assume an XUV absorption efficiency $\epsilon_\mathrm{XUV} = 0.30$. For a lower efficiency of 0.15, which is still consistent with our current understanding of hydrodynamic escape \citep{CHA96,WP13,SHE14}, O$_2$ pressures increase significantly, in many cases by a factor of $\sim 2$. This is because at lower efficiency, hydrogen escape is slower and O$_2$ drag is less efficient, leading to a quicker oxygen buildup. Our calculations are therefore conservative in this sense.

Finally, we neglected tidal heating and orbital evolution due to tides. For late M dwarfs, tidal evolution can be quite strong in the HZ \citep{BAR08,LUG14}, meaning that planets on circular orbits in the HZ today may have started outside of the HZ on eccentric orbits. While this may help reduce the insolation on such planets early on, tidal heating could provide sufficient surface heat flux to trigger a runaway \citep{BAR13}. A tidal runaway could lead to a longer period of water loss and oxygen buildup; this will be investigated in future work.

\section{Conclusions}
\label{sec:conclusions}
We have shown that the extended pre-main sequence contraction phase of M dwarfs causes the habitable zones of these stars to move inwards by up to an order of magnitude in semi-major axis over the course of the first several hundred Myr. Since terrestrial M dwarf planets probably form within $\sim 10$ Myr after the formation of the parent star, many planets currently in the HZs of M dwarfs were not in the HZs when they formed. If these planets formed with water, they may have experienced prolonged runaway greenhouses, lasting between $\sim$ 10 Myr for high mass M dwarfs and $\sim$ 1 Gyr for the lowest mass M dwarfs. Such prolonged runaways could lead to planetary evolution fundamentally different from Earth's, potentially compromising their habitability in the long run.

During a runaway greenhouse, photolysis of water vapor in the stratosphere followed by the hydrodynamic escape of the upper atmosphere can lead to the rapid loss of a planet's surface water. Because hydrogen escapes preferentially over oxygen, large quantities of O$_2$ also build up. We have shown that planets currently in the HZs of M dwarfs may have lost up to several tens of terrestrial oceans (TO) of water during the early runaway phase, accumulating O$_2$ at a constant rate that is set by diffusion: about 5 bars/Myr for Earth-mass planets and 25 bars/Myr for super-Earths. At the end of the runaway phase, this leads to the buildup of hundreds to thousands of bars of O$_2$, which may or may not remain in the atmosphere. We considered two limiting cases regarding the oxygen: (i) highly efficient surface sinks, resulting in an atmospheric O$_2$ content that is always low; and (ii) inefficient O$_2$ sinks, leading to quick atmospheric buildup.

In the first case, we assume that water vapor is the dominant atmospheric species and that atmospheric escape occurs in the energy-limited regime. Both hydrogen and oxygen escape in proportions controlled by the stellar XUV flux. We find that for $M_\star \lesssim 0.3\mathrm{M}_\odot$, nearly all Earth-mass planets in the HZ lose at least 1 TO, though tens of TO are typically lost for $M_\star \lesssim 0.15\mathrm{M}_\odot$. For $0.3\mathrm{M}_\odot \lesssim M_\star \lesssim 0.6\mathrm{M}_\odot$, several TO are lost in the center of the HZ and close to the inner edge. The surfaces of these planets undergo extreme oxidation, absorbing the equivalent of hundreds to thousands of bars of O$_2$.

In the second case, we assume that O$_2$ is produced faster than surface sinks can remove it, resulting in an oxygen-rich atmosphere. We thus calculate escape rates according to the diffusion limit of hydrogen; in this case, the escape flux is insufficient to drag any of the oxygen off to space. Because of the lower escape flux, water loss rates are slightly lower. However, several TO are still lost, particularly around low mass M dwarfs and close to the inner edge of the HZ. Oxygen amounts, on the other hand, are slightly \emph{higher}; planets around all M dwarfs can develop atmospheres with hundreds to thousands of bars of O$_2$.

Perhaps counter-intuitively, we find that both the amount of water lost and the final oxygen pressure scale with planet mass; super-Earths tend to lose substantially more water and develop more massive O$_2$ atmospheres than Earth-mass planets. Despite their higher surface gravity, which reduces the total energy-limited escape rate, super-Earths lose water primarily via the escape of hydrogen, which causes a faster net loss of the oceans compared to the case in which both hydrogen and oxygen escape (which primarily occurs on Earth-mass planets). This is also the case for loss at the diffusion limit, since the escape flux scales with the surface gravity of the planet. We showed that as a result of this enhanced escape, some recently discovered super-Earths in the HZs of M dwarfs such as GJ 667Cc could have lost on the order of 10 TO and built up $\sim$ 2000 bars of O$_2$.

Given the variety of possible planetary compositions and processes capable of removing O$_2$ from the atmosphere, our two cases should roughly bracket the evolution of many exoplanets orbiting M dwarfs. Many of these planets, in particular super-Earths, could retain enough atmospheric O$_2$ to be spectroscopically detectable by future missions such as the James Webb Space Telescope and the WSO-UV space observatory \citep{FOS14}. Our work thus strengthens the results of \cite{WP14}, \cite{TIA14}, and \cite{DOM14} that O$_2$ in a planetary atmosphere is not a reliable biosignature; in fact, such elevated quantities of atmospheric oxygen could potentially be an anti-biosignature. The habitability of many planets around M dwarfs must thus be questioned.

\begin{acknowledgments}
\small{
RL wishes to thank Benjamin Charny, Eddie Schwietermann, Russell Deitrick, Michael Tremmel, and Ty Robinson for helpful discussions about the physics of water loss and for some excellent insight on several matters. RL and RB wish to thank David Catling, James Kasting, Ravi Kopparapu, and Robin Wordsworth for their great feedback and suggestions regarding the runaway greenhouse process, Peter Driscoll for helping with some tough geophysics questions, John Baross for musings on the role of oxygen in the evolution of life, and the rest of the VPL team for their awesome support.

This work was supported by the NASA Astrobiology Institute's Virtual Planet Laboratory under Cooperative Agreement solicitation NNH05ZDA001C and by a generous fellowship from the ARCS Seattle chapter.}

\end{acknowledgments}

\begin{appendix}
\section{A. Rate of Ocean Loss and Oxygen Buildup in the Energy-Limited Regime}
The energy-limited escape rate $\dot{M}_\mathrm{EL}$ is equal to the sum of the upward mass escape rates of hydrogen and oxygen:
\begin{align}
\label{eq:mel}
\dot{M}_\mathrm{EL} = \dot{m}_\mathrm{H}^\uparrow + \dot{m}_\mathrm{O}^\uparrow.
\end{align}
Assuming all of the hydrogen and oxygen comes from photolysis of water, the mass production rate of oxygen is eight times that of hydrogen:
\begin{align}
\label{eq:simple8}
\dot{m}_\mathrm{O} = 8\dot{m}_\mathrm{H}.
\end{align}
Assuming further that all of the hydrogen escapes hydrodynamically, $\dot{m}_\mathrm{H} = \dot{m}_\mathrm{H}^\uparrow$. Since the oxygen either escapes or accumulates in the atmosphere, $\dot{m}_\mathrm{O} = \dot{m}_\mathrm{O}^\uparrow + \dot{m}_\mathrm{O}^\mathrm{atm}$, where $\dot{m}_\mathrm{O}^\mathrm{atm}$ is the rate of oxygen buildup in the atmosphere. We thus have
\begin{align}
\label{eq:moupdown}
\dot{m}_\mathrm{O}^\uparrow + \dot{m}_\mathrm{O}^\mathrm{atm} = 8\dot{m}_\mathrm{H}^\uparrow.
\end{align}
Given that the particle and mass fluxes are related by
\begin{align}
\frac{F_\mathrm{O}}{F_\mathrm{H}} = \frac{1}{16}\frac{\dot{m}_\mathrm{O}^\uparrow}{\dot{m}_\mathrm{H}^\uparrow},
\end{align}
it follows from expression (\ref{eq:oxygenflux}) that
\begin{align}
\label{eq:moupmhup}
\dot{m}_\mathrm{O}^\uparrow = 8\dot{m}_\mathrm{H}^\uparrow\eta,
\end{align}
where $\eta$ is given by (\ref{eq:eta}). Combining expressions (\ref{eq:mel}), (\ref{eq:moupdown}) and (\ref{eq:moupmhup}), we have
\begin{align}
\label{eq:mdothup}
\dot{m}_\mathrm{H}^\uparrow &= \left(\frac{1}{1+8\eta}\right)\dot{M}_\mathrm{EL} \\[0.2cm]
\label{eq:mdotoup}
\dot{m}_\mathrm{O}^\uparrow &= \left(\frac{8\eta}{1+8\eta}\right)\dot{M}_\mathrm{EL} \\[0.2cm]
\label{eq:mdotodown}
\dot{m}_\mathrm{O}^\mathrm{atm} &= \left(\frac{8-8\eta}{1+8\eta}\right)\dot{M}_\mathrm{EL}
\end{align}
with $\dot{M}_\mathrm{EL}$ given by (\ref{eq:dmdt}). Finally, the rate at which the ocean is lost (either to H only escape or H+O escape) is
\begin{align}
\dot{m}_\mathrm{ocean} &= \left(\dot{m}_\mathrm{H}^\uparrow + \dot{m}_\mathrm{O}^\uparrow + \dot{m}_\mathrm{O}^\mathrm{atm} \right) \nonumber\\[0.2cm]
\label{eq:mdotocean}
	&= \left(\frac{9}{1+8\eta}\right)\dot{M}_\mathrm{EL},
\end{align}
which approaches $\dot{M}_\mathrm{EL}$ in the limit $\eta \rightarrow 1$ (H+O escape) and $9\dot{M}_\mathrm{EL}$ in the limit $\eta \rightarrow 0$ (H only escape).\\

\section{B. Dependence on $\mathcal{F}_\mathrm{XUV}$}
Let us now consider expression (\ref{eq:mdotodown}), the rate of oxygen buildup in the atmosphere in the energy-limited regime. Since both $\eta$ and $\dot{M}_\mathrm{EL}$ are functions of $\mathcal{F}_\mathrm{XUV}$, we will rearrange this expression in order to make the dependence more explicit. First, we solve (\ref{eq:eta}) for the XUV flux, making use of (\ref{eq:x}) and (\ref{eq:fhref}):
\begin{align}
\mathcal{F}_\mathrm{XUV} = \left(\frac{40G^2m_\mathrm{H}^2bM_\mathrm{p}^2K_\mathrm{tide}}{kT\epsilon_\mathrm{XUV} R_\mathrm{p}^3}\right) \frac{1+8\eta}{1-\eta},
\end{align}
The energy-limited mass escape rate (\ref{eq:dmdt}) may then be written
\begin{align}
\dot{M}_\mathrm{EL} = \left(\frac{40\pi G m_\mathrm{H}^2 b M_\mathrm{p}}{kT}\right)\frac{1+8\eta}{1-\eta}.
\end{align}
Plugging this into (\ref{eq:mdotodown}), we obtain $\dot{m}_\mathrm{O}^\mathrm{atm}(\eta)$:
\begin{align}
\label{eq:mdotodowneta}
\dot{m}_\mathrm{O}^\mathrm{atm} = \left(\frac{320\pi G m_\mathrm{H}^2 b M_\mathrm{p}}{kT}\right),
\end{align}
which is independent of $\eta$. This means that \emph{the rate of oxygen buildup is constant in time} and does not vary with the XUV flux (provided $\mathcal{F}_\mathrm{XUV} > \mathcal{F}_\mathrm{crit}$). Instead, it depends only on the planet mass and the temperature of the flow. For XUV fluxes below $\mathcal{F}_\mathrm{crit}$, the oxygen escape rate is zero and the rate of buildup in the atmosphere scales linearly with the flux; one must calculate this directly from (\ref{eq:mdotodown}).

When the O$_2$ buildup rate is constant, the expressions for the escape rates of hydrogen (\ref{eq:mdothup}) and oxygen (\ref{eq:mdotoup}) and the expression for the rate of ocean loss (\ref{eq:mdotocean}) greatly simplify. It can be shown from (\ref{eq:mel}) and (\ref{eq:moupdown}) that, provided $\mathcal{F}_\mathrm{XUV} > \mathcal{F}_\mathrm{crit}$,
\begin{align}
\dot{m}_\mathrm{H}^\uparrow &= \frac{1}{9}\dot{M}_\mathrm{EL} + C \\
\dot{m}_\mathrm{O}^\uparrow &= \frac{8}{9}\dot{M}_\mathrm{EL} - C \\
\dot{m}_\mathrm{ocean} &= \dot{M}_\mathrm{EL} + C
\end{align}
where
\begin{align}
C = \frac{\dot{m}_\mathrm{O}^\mathrm{atm}}{9} = \frac{320\pi G m_\mathrm{H}^2 b M_\mathrm{p}}{9kT}.
\end{align}

As we mention in the text, the constant buildup rate of O$_2$ may be understood more easily if we consider the flux of oxygen atoms into the atmosphere. If we divide (\ref{eq:mdotodowneta}) by $4\pi R_\mathrm{p}^2m_\mathrm{H}$, we obtain the rate at which oxygen atoms build up in the atmosphere:
\begin{align}
F_\mathrm{O}^\mathrm{atm} = 5bgm_H/kT,
\end{align}
which is exactly one-half the diffusion limit for hydrogen when $X_\mathrm{O}/X_\mathrm{H} = 1/2$. If we momentarily consider the hydrodynamic flow from the frame of the escaping hydrogen particles, we see that in this case it is the \emph{oxygen} that is diffusing through a static hydrogen background, albeit downwards. The diffusion limit is, from (\ref{eq:difflim}),
\begin{align}
\phi_\mathrm{O} &= \frac{bg(m_\mathrm{H}-m_\mathrm{O})}{kT(1+X_\mathrm{H}/X_\mathrm{O})} \\[0.1in]
								&= -5bgm_\mathrm{H}/kT,
\end{align}
where the negative sign indicates a downward flux. This result is no coincidence; it implies that oxygen is retained \emph{at its diffusion limit}. In other words, in order for the oxygen to accumulate in the atmosphere, it must diffuse out of the hydrodynamic flow that is attempting to carry it away, and the rate at which it can do so is capped at the diffusion limit.

When the escape of hydrogen is diffusion-limited instead of energy-limited, the rate at which oxygen builds up in the atmosphere is again one-half the diffusion limit (since one oxygen atom is left behind for every two escaping hydrogen atoms). Initially, therefore, the rate of buildup is the same as in the energy-limited case. However, as $X_\mathrm{O}/X_\mathrm{H}$ begins to increase, both the H escape rate and the O$_2$ buildup rate decrease. Nonetheless, only once oxygen becomes the dominant species in the atmosphere does the rate at which it is produced begin to taper off.

\begin{figure}[t]
  \begin{center}
    \leavevmode
      \psfig{file=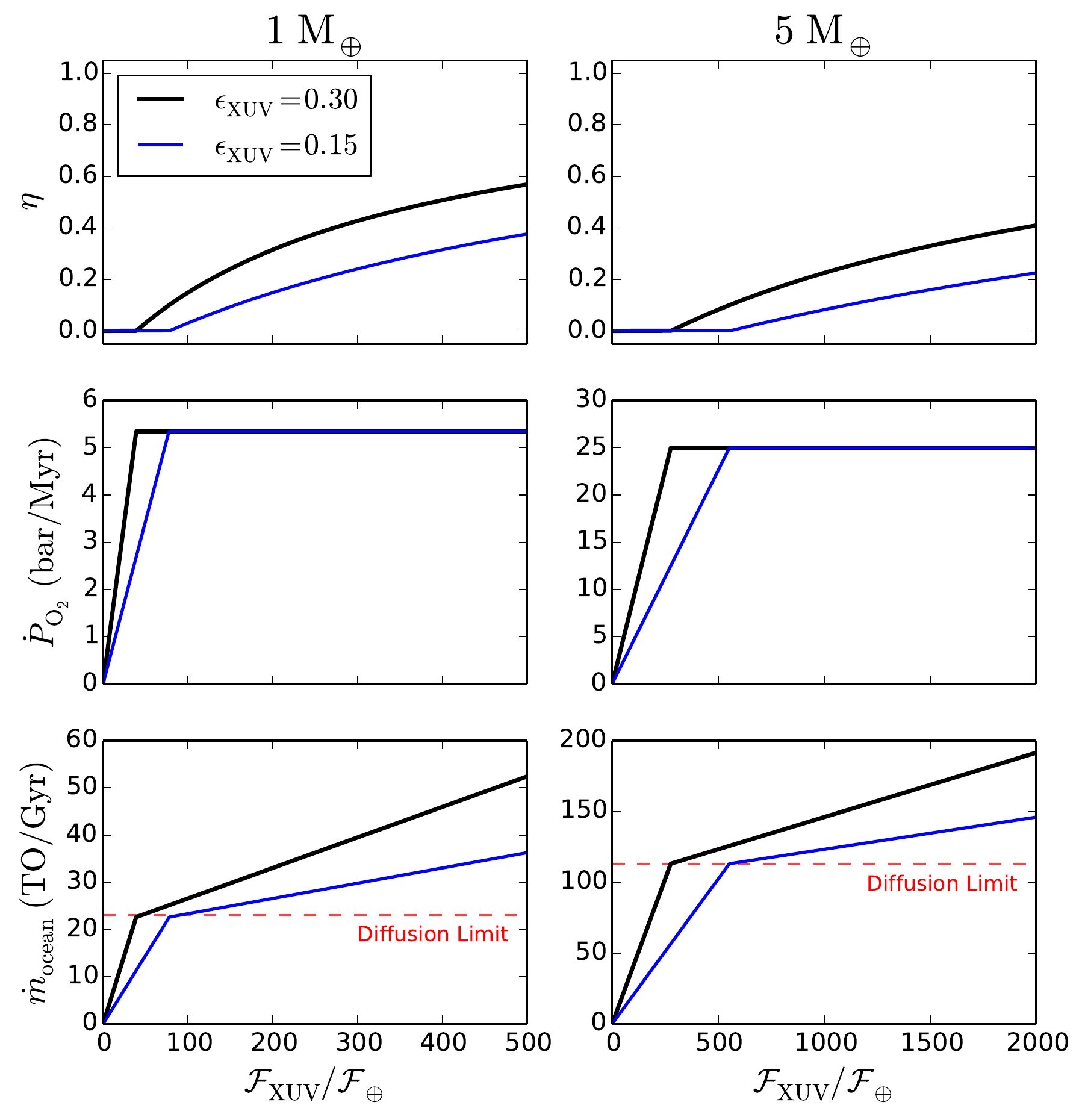,width=4in}
       \caption{Dependence of the oxygen escape parameter $\eta$ (top), the rate of oxygen buildup $\dot{P}_\mathrm{O_2}$ (center), and the ocean loss rate $\dot{m}_\mathrm{ocean}$ (bottom) on the XUV flux for a 1 M$_\oplus$ Earth (left) and a 5 M$_\oplus$ super-Earth (right) in the energy-limited regime. Results for two different XUV efficiencies are plotted: 0.30 (black) and 0.15 (blue). For $\eta = 0$, corresponding to $\mathcal{F}_\mathrm{XUV} < \mathcal{F}_\mathrm{crit}$, the rates of oxygen buildup and ocean loss are linear in $\mathcal{F}_\mathrm{XUV}$. For $\eta > 0$, the oxygen buildup rate is constant at $\sim$ 5 bar/Myr for the Earth and $\sim$ 25 bar/Myr for the super-Earth. The rate of ocean loss is still linear in $\mathcal{F}_\mathrm{XUV}$, but increases more slowly. For reference, in the bottom panel we plot the diffusion-limited ocean loss rate as a dashed red line.}
     \label{fig:eta}
  \end{center}
\end{figure}

For convenience, we now provide expressions for the rate of change of the equivalent oxygen pressure with time, assuming $X_\mathrm{O}/X_\mathrm{H} = 1/2$. Noting that
\begin{align}
\dot{P}_\mathrm{O_2} \approx \frac{GM_\mathrm{p}}{4\pi R_\mathrm{p}^4}\dot{m}_\mathrm{O}^\mathrm{atm},
\end{align}
we obtain the following from (\ref{eq:mdotodowneta}) and (\ref{eq:mdotodown}):
\begin{align}
\label{eq:podot}
\dot{P}_\mathrm{O_2} =
  \begin{dcases}
   5.35 \left( \frac{M}{\mathrm{M_\oplus}} \right)^2	\left( \frac{R}{\mathrm{R_\oplus}} \right)^{-4}\ \mathrm{bars\ Myr^{-1}} & \text{if } \mathcal{F}_\mathrm{XUV} \geq \mathcal{F}_\mathrm{crit} \\
   0.138 \left( \frac{\mathcal{F}_\mathrm{XUV}}{\mathrm{\mathcal{F}_\oplus}} \right) \left( \frac{R}{\mathrm{R_\oplus}} \right)^{-1} \left( \frac{\epsilon_\mathrm{XUV}}{0.30} \right)\ \mathrm{bars\ Myr^{-1}}     	& \text{if } \mathcal{F}_\mathrm{XUV} < \mathcal{F}_\mathrm{crit}.
  \end{dcases}
\end{align}
In Figure~\ref{fig:eta} we plot $\eta$, $\dot{P}_\mathrm{O_2}$, and $\dot{m}_\mathrm{ocean}$ as a function of $\mathcal{F}_\mathrm{XUV}$ for two planet masses and two XUV absorption efficiencies. Note the constant value of $\dot{P}_\mathrm{O_2}$ and the linear behavior of $\dot{m}_\mathrm{ocean}$ above $\mathcal{F}_\mathrm{crit}$.

\end{appendix}

\clearpage

\bibliographystyle{apj}
\bibliography{arxiv}

\end{document}